\documentclass[aps,reprint,superscriptaddress,longbibliography,prb,floatfix]{revtex4-2}
\usepackage{graphicx}
\usepackage{tikz}
\usepackage{amsmath}

\usepackage{amsfonts}
\usepackage{comment}
\usepackage{bbm}
\usepackage{bm}
\usepackage{comment}
\usepackage{esint}
\usepackage{cancel}
\usepackage[dvipsnames]{xcolor}
\usepackage{microtype}
\usepackage[normalem]{ulem}
\usepackage{mathrsfs}
\usepackage{multirow}
\usepackage{array}
\usepackage{lipsum}
\usepackage[unicode=true, colorlinks=true, citecolor={blue!80!black}, urlcolor={blue!50!black}, linkcolor = {blue!80!black}]{hyperref}

\usetikzlibrary{shapes.geometric, arrows.meta, positioning}

\DeclareMathOperator{\tr}{\mathrm{Tr}}

\newcommand{\abs}[1]{\left\vert#1\right\vert}
\newcommand{\ket}[1]{\left\vert#1\right\rangle}
\newcommand{\bra}[1]{\left\langle#1\right\vert}

\newcommand{\av}[1]{\left\langle#1\right\rangle}

\graphicspath{{../figures/}}

\begin{document}
\title{
Dynamics of Majorana tetron qubits under quasiparticle poisoning
}
\author{Sauri Bhattacharyya}
\affiliation{Dipartimento di Fisica, Sapienza Università di Roma, Piazzale Aldo Moro 2, 00185 Rome, Italy}
\author{Bernard van Heck}
\affiliation{Dipartimento di Fisica, Sapienza Università di Roma, Piazzale Aldo Moro 2, 00185 Rome, Italy}

\date{\today}
\begin{abstract}
We study the dissipative dynamics of a Majorana tetron qubit in the presence of extrinsic quasiparticle poisoning due to the coupling to external leads. From the Bloch-Redfield equation describing a finite-size topological superconductor hosting four Majorana zero modes and tunnel-coupled to fermionic reservoirs, we recover analytical expressions for the decoherence rate of a Majorana qubit at arbitrary values of the charging energy. The analysis shows that the exponential suppression of the decoherence rate is gradually removed by the energy splitting of the qubit states. These results can be useful to understand time-domain experimental data in Majorana qubit prototypes.
\end{abstract}

\maketitle


\section{Introduction}

In fault-tolerant quantum computation based on Majorana zero modes~\cite{nayak2008,dassarma2015}, qubits are encoded in the joint fermion parity of well-separated Majorana zero modes in a gapped topological superconductor~\cite{flensberg2021}.
The separation between Majorana zero modes ensures that the computational states are degenerate in energy, up  to corrections $\propto e^{-L/\xi}$ at Majorana distances $L\gg \xi$, the coherence length in the topological phase.
Even then, fermionic excitations may lead to computational errors due to \emph{quasiparticle poisoning}. However, in thermal equilibrium the energy gap $\Delta$ makes the likelihood of their presence exponentially small, $\propto e^{-\Delta/T}$ with $T\ll\Delta$.

Taken together, these two conditions provide the topological protection~\cite{cheng2012,bonderson2013} that underpins the prediction of long coherence times~\cite{knapp2018}. In real devices, it remains necessary to eliminate other forms of quasiparticle poisoning which do not originate from thermally excited quasiparticles in the superconductor. These may originate from external fermionic reservoirs (sometimes referred to as \emph{extrinsic} quasiparticle poisoning~\cite{goldstein2011,budich2012,rainis2012}), as well as from the out-of-equilibrium population of above-gap excitations (\emph{intrinsic} quasiparticle poisoning~\cite{karzig2021,alase2025}).

To mitigate extrinsic quasiparticle poisoning, scalable proposals based on semiconductor-superconductor heterostructures~\cite{aasen2016,karzig2017,aasen2025} rely on the charging energy of a mesoscopic superconducting island~\cite{fu2010}, which assigns a finite energy cost to the addition or removal of a charge $e$ to the island. If we call this energy cost $U$, the quasiparticle poisoning rates should be suppressed by an exponential factor $\propto e^{-U/T}$, restoring the topological protection provided that $T\ll U$.

The scalable logical unit of such proposals is the \emph{tetron}, a superconducting island hosting four Majorana zero modes~\cite{karzig2017} (see also Refs.~\cite{vijay2016,plugge2017,schrade2018}). In a tetron, one bit of quantum information can be encoded in the two degenerate ground states of the island, and all required quantum operations can be executed via projective fermion parity measurements on different pairs of Majorana zero modes.

Early prototypes of tetron devices have been realized experimentally in InAs--Al~\cite{aghaee2023,microsoft2025} and InAs--Pb heterostructures~\cite{aghaee2026}. While the tuning of these devices into a genuine topological phase is made difficult by disorder and remains strongly debated~\cite{dassarma2023,legg2026,microsoft2026reply}, on the more technical side both continuous parity monitoring as well as single-shot parity measurements via dispersive readout mediated by a quantum dot have been reported~\cite{microsoft2025,aghaee2025}. Time-domain fermion parity measurements have also been reported~\cite{vanloo2026,zatelli2026} on minimal Kitaev chains realized with hybrid quantum dot arrays~\cite{sau2012,dvir2023,ten2025}, which can be used to realize tetron-like devices~\cite{pan2025,nitsch2025}.

These developments will hopefully lead to the experimental study of quantum dynamics and decoherence in Majorana-based qubits, whose characterization will precede their use in quantum information processing.
It is therefore a good time to study such devices from an open quantum system perspective~\cite{breuer2002}.

In this work, we provide a quantitative theory of the dynamics of a Majorana tetron qubit subject to extrinsic quasiparticle poisoning.
Although it is well known that all sources of extrinsic quasiparticle poisoning must be minimized in Majorana-based devices, this may not always be possible, especially in current devices that need transport contacts for the tuning of the topological phase.
Our theory provides an explicit derivation of the exponential protection from decoherence of a Majorana qubit as well as, crucially, the departure from topological protection away from the ideal-case scenario.

\section{The model}
\label{sec:model}

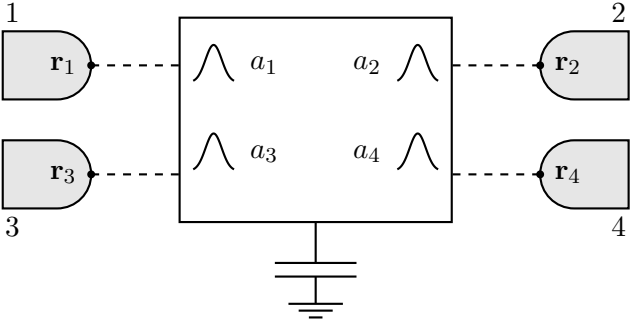
\begin{figure}[t!]
    \centering
    \begin{tikzpicture}[
    scale=0.9,
    every node/.style={font=\fontsize{11pt}{13pt}\selectfont},
    line width=0.8pt,
    port style/.style={fill=lightgray!40, draw=black}
]

    \draw (0,0) rectangle (4,3);

    \draw plot[domain=-0.3:0.3, samples=25, smooth] (\x + 0.5, {0.55*exp(-35*\x*\x) + 2.05});
    \node[anchor=west] at (0.9, 2.3) {$a_1$};

    \draw plot[domain=-0.3:0.3, samples=25, smooth] (\x + 0.5, {0.55*exp(-35*\x*\x) + 0.75});
    \node[anchor=west] at (0.9, 1.0) {$a_3$};

    \node[anchor=east] at (3.1, 2.3) {$a_2$};
    \draw plot[domain=-0.3:0.3, samples=25, smooth] (\x + 3.5, {0.55*exp(-35*\x*\x) + 2.05});

    \node[anchor=east] at (3.1, 1.0) {$a_4$};
    \draw plot[domain=-0.3:0.3, samples=25, smooth] (\x + 3.5, {0.55*exp(-35*\x*\x) + 0.75});

    \node[anchor=south west] at (-2.7, 2.8) {$1$};
    \draw[port style] (-2.6, 2.8) -- (-1.8, 2.8) arc (90:-90:0.5 and 0.5) -- (-2.6, 1.8) -- cycle;
    \filldraw (-1.3, 2.3) circle (1.2pt) node[left=2pt] {$\mathbf{r}_1$};
    \draw[dashed] (-1.3, 2.3) -- (0, 2.3);

    \node[anchor=north west] at (-2.7, 0.2) {$3$};
    \draw[port style] (-2.6, 1.2) -- (-1.8, 1.2) arc (90:-90:0.5 and 0.5) -- (-2.6, 0.2) -- cycle;
    \filldraw (-1.3, 0.7) circle (1.2pt) node[left=2pt] {$\mathbf{r}_3$};
    \draw[dashed] (-1.3, 0.7) -- (0, 0.7);

    \node[anchor=south east] at (6.7, 2.8) {$2$};
    \draw[port style] (6.6, 2.8) -- (5.8, 2.8) arc (90:270:0.5 and 0.5) -- (6.6, 1.8) -- cycle;
    \filldraw (5.3, 2.3) circle (1.2pt) node[right=2pt] {$\mathbf{r}_2$};
    \draw[dashed] (5.3, 2.3) -- (4, 2.3);

    \node[anchor=north east] at (6.7, 0.2) {$4$};
    \draw[port style] (6.6, 1.2) -- (5.8, 1.2) arc (90:270:0.5 and 0.5) -- (6.6, 0.2) -- cycle;
    \filldraw (5.3, 0.7) circle (1.2pt) node[right=2pt] {$\mathbf{r}_4$};
    \draw[dashed] (5.3, 0.7) -- (4, 0.7);

    \draw (2, 0) -- (2, -0.6);
    \draw (1.4, -0.6) -- (2.6, -0.6);
    \draw (1.4, -0.8) -- (2.6, -0.8);
    \draw (2, -0.8) -- (2, -1.2);
    \draw (1.6, -1.2) -- (2.4, -1.2);
    \draw (1.75, -1.3) -- (2.25, -1.3);
    \draw (1.9, -1.4) -- (2.1, -1.4);

\end{tikzpicture}
    \caption{A schematic diagram of a tetron device: a floating superconducting island (white box) contains four localized Majorana zero modes $a_1, a_2, a_3, a_4$. The superconductor is tunnel-coupled (dashed lines) to four fermionic leads (grey). }
    \label{fig:setup}
\end{figure}

As depicted in Fig.~\ref{fig:setup}, a tetron device consists of a topological superconducting island hosting four Majorana zero modes.
We imagine that the four Majorana zero modes are coupled to fermionic leads which act as an external source of quasiparticles.
In experimental setups, they may serve the purpose of transport probes for the topological superconductor, or be part of the measurement apparatus for fermion parity measurements on the tetron.

The four well-separated Majorana zero modes are represented by Hermitian operators $a_i$ obeying
\begin{equation}
\{a_i, a_j\}=\delta_{ij}\,,
\end{equation}
for $i, j=1,\dots, 4$. Because these modes are the local degrees of freedom non-locally encoding a qubit, we shall refer to them as the computational Majorana modes.

\subsection{Hamiltonian}

We adopt the following effective Hamiltonian to describe the quantum dynamics of the computational Majorana zero modes:
\begin{equation}\label{eq:microscopic_model}
H = H_A + H_{AB} + H_B
\end{equation}
The first term describes the tetron in isolation from the leads:
\begin{equation}
H_A = \frac{i}{2} \sum_{ij} A_{ij}a_ia_j + 2U a_{1}a_{2}a_{3}a_{4}\,.\\
\end{equation}
The quadratic part of $H_A$ contains the direct couplings among the computational Majoranas, gathered in the $4\times 4$ matrix $A$, which by construction is real and antisymmetric ($A_{ij}=-A_{ji}$).
The direct couplings are non-zero due to the finite spatial extent of the zero modes and the finite size of the tetron.
The interaction term, which has a strength $U$, ensues from the finite capacitance to ground of the superconducting island hosting the Majoranas. It is proportional to the total fermion parity of the Majorana zero modes,
\begin{equation}
P=-4a_1a_2a_3a_4\,,
\end{equation}
which is equal to $P=+1$ for even-parity states and $P=-1$ for odd-parity states.

The computational Majorana zero modes are coupled to four fermionic leads by single particle tunneling. For simplicity we adopt a point contact model, in which each of the four reservoirs contacts the tetron at a single position $\mathbf{r}_i$. This is described by the second term $H_{AB}$:
\begin{equation}
H_{AB} = \sum_{ij} w_{ij}\,a_i\,c(\mathbf{r}_j) +\textrm{h.c.}\,.
\end{equation}
where $w_{ij}$ is a complex tunneling amplitude and $c(\mathbf{r}_j)$ annihilates a fermion in the $j$-th reservoir at position $\mathbf{r}_j$ (see Fig.~\ref{fig:setup}).
It is useful to write the tunneling Hamiltonian in the form $H_{AB}=-i \sum_i a_i B_i$, with Hermitian channel operators
\begin{equation}
B_i = i\sum_j w_{ij} c(\mathbf{r}_j)-w_{ij}^* c^\dagger(\mathbf{r}_j)\,.
\end{equation}
Note that this Hamiltonian allows a single Majorana zero mode to couple to multiple leads. We will often, however, invoke a simplifying assumption that $a_i$ couples only to lead $i$, and that the coupling strengths are all equal. This corresponds to the choice
\begin{equation}\label{eq:simple_symmetric_couplings}
w_{ij}=w\,\delta_{ij}
\end{equation}
for some complex amplitude $w$, so that the channel operator is reduced to the simpler form $B_i = w\,c(\mathbf{r}_i)-w^*\,c^\dagger(\mathbf{r}_j)$. In Sec.~\ref{sec:results_andreev}, we consider a scenario in which this assumption is relaxed.

The third term $H_{B}$ is the Hamiltonian of the leads, which are assumed to be gapless and in thermal equilibrium at a temperature $T$.
Rather than the specific form of $H_B$, it is useful to present the correlation function relevant for the open system dynamics, which is
\begin{equation}\label{eq:Cij}
C_{ij}(t) = \frac{1}{Z}\,\tr_B[B_i(t) B_j(0) e^{-\beta H_B}]\,.
\end{equation}
with $B_i(t)=e^{iH_Bt}B_ie^{-iH_Bt}$, $Z=\tr_B e^{-\beta H_B}$, and $\beta=1/T$. By the cyclicity of the trace, the equilibrium correlation function obeys the Kubo-Martin-Schwinger condition
\begin{equation}\label{eq:KMS}
C_{ij}(t)=C_{ji}(-t-i\beta)\,,
\end{equation}
as well as $C_{ji}(t)=C_{ij}^*(-t)$. If the couplings take the simple form of Eq.~\eqref{eq:simple_symmetric_couplings}, and if the leads are themselves assumed to be all identical, the correlation function factorizes as
\begin{equation}
C_{ij}(t)=\delta_{ij}\,C(t)\,,
\end{equation}
and we only have a single function $C(t)$ to characterize. The bath correlation function is studied in more detail in Appendices~\ref{app:correlation_functions} and~\ref{app:wide-band}.

\subsection{Time and energy scales}

In general, the behavior of $C(t)$ is controlled by two time scales: one pertaining to the bath and one to the system-bath coupling strength $w$. The first is the 
bath correlation time $\tau_B$, formally defined as~\cite{mozgunov2020,nathan2020}
\begin{equation}\label{eq:tauB}
\tau_B = \frac{\int_0^\infty t \abs{C(t)}dt}{\int_0^\infty \abs{C(t)}dt}
\end{equation}
For a wide-band fermionic reservoir in thermal equilibrium, as shown in Appendix~\ref{app:wide-band}, the bath correlation time scale is controlled by temperature:
\begin{equation}
\tau^{-1}_B \approx \pi T\,.
\end{equation}
Importantly, $\tau_B$, is independent from the system-bath coupling strength $w$. The latter determines instead a bare timescale for the bath-induced decoherence of the system, which we indicate as $\tau_D$:
\begin{equation}\label{eq:tauD}
\tau_D^{-1} = \int_0^\infty \abs{C(t)}dt\,.
\end{equation}
Again in Appendix~\ref{app:wide-band}, we show that this time scale is controlled by the tunneling rate between the system and the reservoir:
\begin{equation}\label{eq:Gamma0-definition}
\tau_D^{-1} \approx \Gamma_0\,,\quad \Gamma_0 = 2\pi \abs{w}^2 \nu(0)
\end{equation}
with $\nu(0)$ the density of states in the leads at the Fermi level.

In Sec.~\ref{sec:bloch-redfield} we derive the master equation for the system under the assumption of weak system-bath coupling, which allows us to employ the Born-Markov approximation, valid under the condition
\begin{equation}
\tau_D \gg \tau_B\,.
\end{equation}
For our fermionic reservoirs, this is equivalent to
\begin{equation}
\Gamma_0 \ll T\,.
\end{equation}

Two additional energy scales determine the dynamics of the tetron in isolation, which is discussed in detail in Section~\ref{sec:unitary_dynamics}: the interaction $U$ and the magnitude of the overlap couplings $A_{ij}$, which we can characterize by the energy scale
\begin{equation}
\epsilon = \max |A_{ij}|\,.
\end{equation}
The regime of topological protection of the Majorana qubit corresponds to the following hierarchy of time scales,
\begin{equation}
\epsilon \ll \Gamma_0 \ll T \ll U\,.
\end{equation}
We refer to this regime as the \emph{degenerate} regime. It is studied in Sec.~\ref{sec:degenerate}, which covers the open system dynamics under the assumption $\epsilon \ll \Gamma_0$.

The regime of topological protection is gradually removed upon increasing $\epsilon$. In Sec.~\ref{sec:secular} we study the master equation in the regime $\epsilon \gg \Gamma_0$, which we refer to as the \emph{secular} regime. We then show that the validity of the results can be extended across intermediate values $\epsilon \sim \Gamma_0$, all the way to the degenerate limit. The results of both Sections~\ref{sec:degenerate} and~\ref{sec:secular} apply to arbitrary ratios $U/T$, always provided that $T\gg \Gamma_0$.

\section{Unitary dynamics of the tetron}
\label{sec:unitary_dynamics}

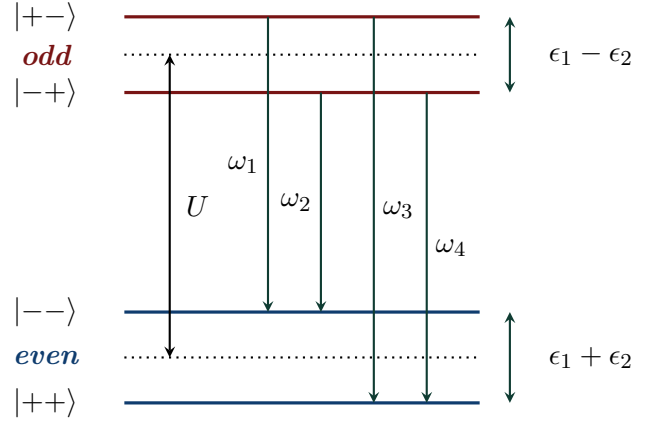
\begin{figure}[t!]
    \centering
    \begin{tikzpicture}[>=stealth, scale=1]
    \definecolor{topred}{RGB}{120, 20, 20}
    \definecolor{botblue}{RGB}{15, 60, 110}
    \definecolor{darkgreen}{RGB}{15, 60, 50}

    \tikzset{
        text font/.style={font=\fontsize{11}{13}\selectfont, text=black}
    }

    \draw[very thick, topred] (0.5, 3.6) -- (5.2, 3.6);
    \draw[very thick, topred] (0.5, 2.6) -- (5.2, 2.6);
    \draw[thick, dotted] (0.5, 3.1) -- (5.2, 3.1);
    
    \node[left, text font] (k_plus_minus) at (0, 3.6) {$\ket{+-}$};
    \node[topred, font=\bfseries\itshape\fontsize{11}{13}\selectfont] at (k_plus_minus.center |- 0, 3.1) {odd};
    \node[left, text font] at (0, 2.6) {$\ket{-+}$};

    \draw[very thick, botblue] (0.5, -0.3) -- (5.2, -0.3);
    \draw[very thick, botblue] (0.5, -1.5) -- (5.2, -1.5);
    \draw[thick, dotted] (0.5, -0.9) -- (5.2, -0.9);
    
    \node[left, text font] (k_minus_minus) at (0, -0.3) {$\ket{--}$};
    \node[botblue, font=\bfseries\itshape\fontsize{11}{13}\selectfont] at (k_minus_minus.center |- 0, -0.9) {even};
    \node[left, text font] at (0, -1.5) {$\ket{++}$};

    \draw[<->, thick, black] (1.1, -0.9) -- (1.1, 3.1);
    \node[right, text font, font=\fontsize{11}{13}\selectfont\itshape] at (1.2, 1.1) {$U$};

    \draw[<->, thick, darkgreen] (5.6, 2.6) -- (5.6, 3.6);
    \node[right, text font] at (6.0, 3.1) {$\epsilon_1 - \epsilon_2$};

    \draw[<->, thick, darkgreen] (5.6, -1.5) -- (5.6, -0.3);
    \node[right, text font] at (6.0, -0.9) {$\epsilon_1 + \epsilon_2$};

    \draw[->, thick, darkgreen] (2.4, 3.6) -- (2.4, -0.3) 
        node[midway, left, text font] {$\omega_1$};
        
    \draw[->, thick, darkgreen] (3.1, 2.6) -- (3.1, -0.3) 
        node[midway, left, text font] {$\omega_2$};
        
    \draw[->, thick, darkgreen] (3.8, 3.6) -- (3.8, -1.5) 
        node[midway, right, text font] {$\omega_3$};
        
    \draw[->, thick, darkgreen] (4.5, 2.6) -- (4.5, -1.5) 
        node[midway, right, text font] {$\omega_4$};

\end{tikzpicture}
    \caption{The energy-level scheme of the isolated Majorana system. It comprises of a low-energy even-parity doublet, and a high-energy odd-parity one. The intra-doublet splittings are due to the direct couplings among the zero modes, while the inter-doublet gap is due to interactions. The dotted lines show the transitions corresponding to the four positive Bohr frequencies of Eqs.~\ref{eq:Bohr-freqs}.}
    \label{fig:level_scheme}
\end{figure}

Before analyzing the open dynamics of the tetron, it is useful to examine the unitary dynamics under the isolated system Hamiltonian $H_{A}$. The spectrum of $H_A$ is obtained by using an orthogonal matrix $O$ to put the matrix $A$ into canonical form- 
\begin{equation}\label{eq:standard_form_bath_hamiltonian}
O A O^T = \begin{bmatrix}
    0 & \epsilon_1 & 0 & 0 \\ -\epsilon_1 & 0 & 0 & 0 \\
    0 & 0 & 0 & \epsilon_2 \\ 0 & 0 & -\epsilon_2 & 0
\end{bmatrix}\,.
\end{equation}
Without loss of generality, we assume $\epsilon_1 > \epsilon_2 > 0$. In terms of the operators $\tilde{a}=Oa$,
one has
\begin{equation}
H_{A} = i\epsilon_1\tilde{a}_1\tilde{a}_2+i\epsilon_2\tilde{a}_3\tilde{a}_4 + 2U \tilde{a}_1\tilde{a}_2\tilde{a}_3\tilde{a}_4\,.\\
\end{equation}
The three operators in $H_A$ commute with each other and so the spectrum is easily constructed. It is convenient to introduce joint parity operators in the eigenbasis,
\begin{equation}
\tilde{P}_{ij}=-2i\tilde{a}_i\tilde{a}_j\,,
\end{equation}
which have eigenvalues $\pm 1$, depending on whether a fermionic mode $(\tilde{a}_i+i\tilde{a}_j)/\sqrt{2}$ is empty or occupied. Note that $P=\tilde{P}_{12}\tilde{P}_{34}$, so that
\begin{equation}
H_A = -\tfrac{1}{2}\epsilon_1 \tilde{P}_{12} -\tfrac{1}{2}\epsilon_2 \tilde{P}_{34}-\tfrac{1}{2}U\tilde{P}_{12}\tilde{P}_{34}\,.
\end{equation}
We can therefore label the eigenstates of $H_A$ as $\ket{\sigma\tau}$ with $\sigma=\pm 1$ the eigenvalues of $\tilde{P}_{12}$ and $\tau=\pm 1$ the eigenvalue of $\tilde{P}_{34}$. Their energies are (see Fig.~\ref{fig:level_scheme})
\begin{equation}
E_{\sigma\tau}=-\tfrac{1}{2}(\sigma \epsilon_1 + \tau \epsilon_2) -\tfrac{1}{2}U\sigma\tau\,.
\end{equation}
and their total parity is $P=\sigma\tau$.
The even and odd sectors of the spectrum are separated by a gap $U$, which is the origin of the protection from quasiparticle poisoning.
We assume that $U>0$ so that the ground state has even parity.

The unitary time evolution of parity observables under $H_A$ is easy to compute. Working in the Heisenberg picture, we obtain 
\begin{subequations}\label{eq:bilinear_averages_ideal}
\begin{align}
\tilde{P}_{12}(t) &= \tilde{P}_{12}(0)\,,\\
\tilde{P}_{13}(t) &= \tilde{P}_{13}(0)\cos[(\epsilon_1+\epsilon_2)t] \\\nonumber
&\qquad+ \tilde{P}_{23}(0)\sin[(\epsilon_1+\epsilon_2)t]\,,\\
\tilde{P}_{14}(t) &= \tilde{P}_{14}(0)\cos[(\epsilon_1-\epsilon_2)t] \\\nonumber
&\qquad + \tilde{P}_{24}(0)\sin[(\epsilon_1-\epsilon_2)t]\,.
\end{align}
\end{subequations}
The time evolution of the remaining parity operators under $H_A$ follows from the conservation of the total parity $P$. Indeed, the conservation of $P$ gives the operator identities
\begin{equation}\label{eq:parity_constraint}
\tilde{P}_{34} = P \tilde{P}_{12}\,,\quad \tilde{P}_{24}=-P\tilde{P}_{13}\,,\quad \tilde{P}_{23}=P\tilde{P}_{14}\,.
\end{equation}
The physical interpretation of  Eqs.~\eqref{eq:bilinear_averages_ideal} and~\eqref{eq:parity_constraint} is the following:  $\tilde{P}_{12}$ and $\tilde{P}_{34}$ are constant-in-time populations of the fermionic modes $c_1=(\tilde{a}_1+i\tilde{a}_2)/\sqrt{2}$ and $c_2=(\tilde{a}_3+i\tilde{a}_4)/\sqrt{2}$, whereas
the other parity operators are precessing coherences between eigenstates of the same total parity.

In discussing dynamics, we may be interested in the time evolution of joint parity operators of \emph{computational} Majorana zero modes,
\begin{equation}
P_{ij}=-2ia_ia_j\,,
\end{equation}
rather than in that of the eigenmodes. Indeed, ideally it is the $P_{ij}$ which are the observables accessible by parity measurement in a tetron device. Using the orthogonal transformation $O$, one obtains
\begin{equation}
P_{ij}(t) = \sum_{ij}O^T_{ii'}O^T_{jj'}\tilde{P}_{i'j'}(t)\,.
\label{eq:bilinear_averages_isolated}
\end{equation}

From this list of basic results, we may collect the following useful observations about the unitary dynamics of parity observables in the tetron.
First, the energy gap $U$ does not enter the time evolution of $P_{ij}(t)$.
Second, joint parities of computational Majorana zero modes in general undergo precession at a frequency that depends on the total parity $P$.
Third, in terms of computational Majorana zero modes it is not always possible to make a distinction between populations and coherences.
This is only possible if the degeneracy is broken and the eigenmodes coincide with the computational ones.

A qubit may be encoded in the two even-parity eigenstates, $\ket{++}$ and $\ket{--}$. In the topologically protected regime, as already mentioned, $\epsilon_1$ and $\epsilon_2$ are exponentially small.
In the ideal limit in which they vanish, the qubit states are degenerate, there is no precession dynamics and no distinction between the computational modes and the eigenmodes. Furthermore, the conservation of total parity in the closed system prevents qubit leakage to the odd-parity sector formed by $\ket{+-}$ and $\ket{-+}$.

Upon introducing the coupling to the external leads, the total parity of the system is no longer conserved, as the tunneling terms in $H_{AB}$ can change it. 
This introduces a leakage timescale out of the qubit subspace.
The coupling to the external leads will also be a source of decoherence within the qubit subspace, that must be added to any dephasing that may occur due to noise in $\epsilon_1$ and $\epsilon_2$.
To determine the timescales associated with leakage and decoherence, we must move on to the study of the open quantum system dynamics.

\section{Bloch-Redfield equation}
\label{sec:bloch-redfield}

Single-particle tunneling between the superconducting island forming the tetron and the reservoirs causes the total parity to change.
There are four possible transitions from an initial state of odd parity to a final state of even parity. They have the following Bohr transition frequencies (see Fig.~\ref{fig:level_scheme}):
\begin{subequations}\label{eq:Bohr-freqs}
\begin{align}
\ket{+-}\to\ket{--}\,:\quad\omega_1 = E_{+-}-E_{--}=U-\epsilon_1\,,\\
\ket{-+}\to\ket{--}\,:\quad\omega_2 = E_{-+}-E_{--}=U-\epsilon_2\,,\\
\ket{+-}\to\ket{++}\,:\quad\omega_3 = E_{+-}-E_{++}=U+\epsilon_2\,,\\
\ket{-+}\to\ket{++}\,:\quad\omega_4 = E_{-+}-E_{++}=U+\epsilon_1\,.
\end{align}
\end{subequations}
Note that the transition frequencies $\omega_s$ are defined as the energy of the initial state minus that of the final state.
Reversed transitions -- from an even parity state to an odd parity state -- have opposite transition frequency.
With this convention, a positive transition frequency corresponds to the release of energy from the tetron into the bath, and thus as an absorption process from the point of view of the reservoir.
On the other hand, a negative transition frequency corresponds to the emission of energy from the bath into the tetron.

The odd-to-even transition frequencies $\omega_s$ are all positive if $U>\epsilon_1>\epsilon_2$, which is the case depicted in Fig.~\ref{fig:level_scheme}.
If $U$ becomes smaller than $\epsilon_1$, then the energy spectrum does not separate anymore into two parity-doublets, but it alternates between even and odd energy levels. In this case, $\omega_1$ becomes negative.
If $U$ is smaller than both $\epsilon_1$ and $\epsilon_2$, then the odd doublet is sandwiched between the two even levels, and both $\omega_1$ and $\omega_2$ are negative.

In general, each of these transitions may be activated by tunneling into or out of any of the computational Majorana zero modes.
Such a process may be described by a jump operator 
\begin{equation}\label{eq:jump_operators}
A_{i,s} = \Pi_{\sigma'_s\tau'_s}\,a_i\,\Pi_{\sigma_s\tau_s}\,,\quad(s=1,2,3,4)\,,
\end{equation}
where $\Pi_{\sigma\tau}$ is a projector onto the eigenstate $\ket{\sigma\tau}$,
\begin{equation}\label{eq:projectors}
\Pi_{\sigma\tau} = \ket{\sigma\tau}\bra{\sigma\tau}=\tfrac{1}{4}(1+\sigma\tilde{P}_{12})\,(1+\tau\tilde{P}_{34})\,,
\end{equation}
and $\{\sigma_s, \tau_s\}$, $\{\sigma'_s, \tau'_s\}$ are the quantum numbers of the initial and final states involved in the transition $s$.
In total, there are sixteen jump operators $A_{i,}$, each of them a linear combination of terms with one or three $a_i$'s.

Using these jump operators, it is possible to formulate a local-in-time master equation for the evolution of the density matrix of the computational Majorana zero modes, $\rho(t)$.
In doing so, we assume that the initial reduced density matrix $\rho(0)$ commutes with the total parity $P$,
\begin{equation}\label{eq:superselection_rule}
[\rho(0),P]=0\,,
\end{equation}
so that it does not contain coherences between the even-parity and odd-parity sectors: $\bra{\sigma\tau}\rho(0)\ket{\sigma'\tau'}=0$ if $\ket{\sigma\tau}$ and $\ket{\sigma'\tau'}$ are states of opposite parity. 
Because, as already discussed, the coherent evolution preserves the total parity, once true at $t=0$ this property remains true at all subsequent times: $[\rho(t),P]=0$.
We can therefore always decompose the density matric into an even and an odd part: $\rho=\rho_\textrm{e}+\rho_\textrm{o}$, with
\begin{equation}
\rho_\textrm{e} = \tfrac{1}{2}(1+P)\rho\quad\textrm{and}\quad\rho_\textrm{o} = \tfrac{1}{2}(1-P)\,,
\end{equation}
which will sometimes come handy.
Note that Eq.~\eqref{eq:superselection_rule} does not prevent population transfer from the even to the odd sector or viceversa, which can and indeed does happen.

Working in the weak coupling limit where $\tau_D\gg \tau_B$, a local-in-time master equation can be derived by invoking the Born-Markov approximation, within a standard open quantum systems approach~\cite{breuer2002}. The detailed derivation is presented in Appendix~\ref{app:wide-band} and yields the following Bloch-Redfield master equation for $\rho(t)$ in the interaction picture:
\begin{widetext}
\begin{equation}\label{eq:Bloch-Redfield}
\frac{d\rho}{dt} = -\frac{1}{4} \sum_{i,j}\sum_{s,s'} \left[e^{i\delta_{ss'}t}\,\Gamma_{ij}(\omega_s) (A^{\dagger}_{i,s'}A_{j,s}\rho - A_{j,s}\, \rho\, A^{\dagger}_{i,s'}) + e^{-i\delta_{ss'}t}\,\Gamma_{ji}(-\omega_s) (A_{i,s'}A^{\dagger}_{j,s}\rho - A^{\dagger}_{j,s}\, \rho\, A_{i,s'}) + \textrm{h.c.}\right]\,,
\end{equation}
\end{widetext}
with $\delta_{ss'}=\omega_{s'}-\omega_s$
and
\begin{equation}
\Gamma_{ij}(\omega) = \int_{0}^{\infty} dt \exp(i\omega t) C_{ij}(t)
\end{equation}
the one-sided Fourier transform of the correlation function.
As customary~\cite{breuer2002}, we split it into its Hermitian and anti-Hermitian parts,
\begin{equation}
\Gamma_{ij}(\omega) = \tfrac{1}{2}\gamma_{ij}(\omega) + i S_{ij}(\omega)
\end{equation}
with
\begin{align}
\gamma_{ij}(\omega) &= \Gamma_{ij}(\omega) + \Gamma_{ji}^{*}(\omega)\,,\\
S_{ij}(\omega) &= [\Gamma_{ij}(\omega)-\Gamma_{ji}^{*}(\omega)]/2i\,. 
\end{align}
The anti-Hermitian part of the correlation function, $S_{ij}(\omega)$, gives rise to the Lamb shift corrections to the unitary dynamics of the system.
The Lamb shift terms arising from $S_{ij}(\omega)$ commute by construction with the system Hamiltonian $H_A$: their effect is only to shift the energy levels $\epsilon_1$ and $\epsilon_2$, and the corresponding precession frequencies in Eqs.~\eqref{eq:bilinear_averages_ideal}. As we are interested in the dissipative effects induced by the reservoirs, we neglect the Lamb shift terms from now own.

We therefore focus on the Hermitian part $\gamma_{ij}(\omega)$, which is the Fourier transform of $C_{ij}(t)$,
\begin{equation}\label{eq:gamma}
\gamma_{ij}(\omega) = \int_{-\infty}^\infty e^{i\omega t} C_{ij}(t)\,dt\,.
\end{equation}
In virtue of the Kubo-Martin-Schwinger condition~\eqref{eq:KMS}, it satisfies detailed balance, namely
\begin{equation}\label{eq:detailedbalance}
\gamma_{ji}(-\omega)=e^{-\beta\omega}\gamma_{ij}(\omega)\,.
\end{equation}
Furthermore, at fixed frequency $\omega$, the matrix $\gamma_{ij}(\omega)$ is positive.

If one makes the simplifying assumption that $w_{ij}=w\,\delta_{ij}$ [Eq.~\eqref{eq:simple_symmetric_couplings}], then $\gamma_{ij}(\omega)=\gamma(\omega)\delta_{ij}$.
In Appendix~\ref{app:correlation_functions}, we show that
\begin{equation}
\gamma(\omega) = 2\pi \abs{w}^2\,\frac{\nu(\omega)+\nu(-\omega)}{1+e^{-\beta\omega}}\,.
\end{equation}
where $\nu(\epsilon)$ is the density of states in the leads. If one can approximate $\nu(\omega)\approx \nu(0)$, which is likely to be a good approximation for a wide-band lead given that the transition frequencies are all of order $U$, then
\begin{equation}\label{eq:gamma_simple}
\gamma(\omega) = \frac{2\Gamma_0}{1+e^{-\beta\omega}}\,,
\end{equation}
with the tunneling rate $\Gamma_0$ defined in Eq.~\eqref{eq:Gamma0-definition}. We will often invoke this simple version of $\gamma(\omega)$ to obtain concrete formulae.

The Bloch--Redfield equation~\eqref{eq:Bloch-Redfield} is not in the Gorini-Kossakowski-Sudarshan-Lindblad (GKSL) form and thus it is not guaranteed to preserve the positivity of the density matrix.
This is due to the presence of the oscillatory factors $\exp(i\delta_{ss'}t)$, which couple transitions associated with different Bohr frequencies.
In our model, the frequency differences take the following possible values:
\begin{subequations}\label{eq:transition_diffs}
\begin{align}
\delta_{11} &= \delta_{22} =\delta_{33} =\delta_{44} =0\,,\\
\delta_{12}&=\delta_{34}=-\delta_{21}=-\delta_{43}= \epsilon_1-\epsilon_2\,,\\
\delta_{13}&=\delta_{24}=-\delta_{31}=-\delta_{42}=\epsilon_1+\epsilon_2\,,\\
\delta_{14}&=-\delta_{41}=2\epsilon_1\,,\\
\delta_{23}&=-\delta_{32}=2\epsilon_2\,.
\end{align}
\end{subequations}
We note that the frequency differences \emph{do not} contain $U$, which is the scale of the gap between the two parity sectors.
This is owing to the fact that the allowed processes flip the total parity successively in opposite directions.

Making further progress on the solution of the Bloch-Redfield equation requires some assumption on $\delta_{ss'}$, and thus essentially $\epsilon$, compared to the other energy scales.
Referring to the nomenclature introduced in the previous Section~\ref{sec:model}, we distinguish between the \emph{degenerate regime} in which all the phase factors can be well approximated with unity over the entire time evolution, and a \emph{secular regime} in which the phase factors average out unless $\delta_{ss'}=0$.
The degenerate regime holds when the direct couplings between Majorana zero modes are very small, $\epsilon \ll \Gamma_0$, and so it includes with the regime of topological protection.
The secular regime, instead, becomes relevant when $\epsilon\gg\Gamma_0$.

\section{The degenerate regime}
\label{sec:degenerate}

\subsection{The ideal case: perfect degeneracy}
\label{sec:degenerate_ideal}

In the ideal scenario behind the design of the tetron, the computational Majorana zero modes are so well separated that all the couplings $A_{ij}$ in the Hamiltonian $H_A$ can be set to zero, so that $H_A=-(U/2)\,P$. Then, the spectrum consists of doubly-degenerate even and odd energy levels ($\epsilon_1=\epsilon_2=0$), separated by an energy gap $U$. With this spectral information as input, and in the simple case in which $\Gamma_{ij}(\omega)=\delta_{ij}\Gamma(\omega)$ [Eq.~\eqref{eq:simple_symmetric_couplings}], the Bloch-Redfield equation~\eqref{eq:Bloch-Redfield} simplifies to
\begin{align}
\label{eq:BR_ideal}
\frac{d}{dt}{\rho}=-\frac{1}{4}\sum_{i,s,s'}&\left[\Gamma(U)(A^{\dagger}_{i,s'}A_{i,s}\rho - A_{i,s}\, \rho\, A^{\dagger}_{i,s'})+\right.\\\nonumber
&+ \left.\Gamma(-U)(A_{i,s'}A^{\dagger}_{i,s}\rho - A^{\dagger}_{i,s}\, \rho\, A_{i,s'})+\textrm{h.c.}\right]\,,
\end{align}
With all transition frequencies degenerate, the sums over $s$ and $s'$ can be performed noting that
\begin{equation}
\sum_s A_{is}=\frac{1}{2}\,a_i\,(1-P)\,.
\end{equation}
After the substitution $\Gamma(U)=\tfrac{1}{2}\gamma(U)+iS(U)$ and a bit of algebra (which, in fact, causes all terms proportional to $S(U)$ to drop out), one arrives at
\begin{equation}
\frac{d\rho}{dt} = \mathcal{D}_0[\rho]
\end{equation}
with a dissipator in the GKSL form,
\begin{align}\nonumber
\mathcal{D}_0[\rho]=&-\frac{\gamma(U)}{2}\bigg[\rho_\textrm{o} -\frac{1}{2}\sum_ia_i\,\rho_\textrm{o}\,a_i\bigg]\\
&-\frac{\gamma(-U)}{2}\bigg[\rho_\textrm{e} -\frac{1}{2}\sum_ia_i\,\rho_\textrm{e}\,a_i\bigg]\,,
\label{eq:dissipator_degenerate}
\end{align}

Due to the degeneracy of energy levels in the even and odd sectors there is no preferred basis in which to solve the master equation.
Nevertheless, to make contact with the calculations in following sections, it is convenient to work in the basis $\ket{\sigma\tau}$ of common eigenstates of $P_{12}$ and $P_{34}$, with $\sigma=\pm$ now the eigenvalue of $P_{12}$ and $\tau=\pm$ the eigenvalue of $P_{34}$.

An important feature of the dynamics is that the diagonal and off-diagonal elements of the reduced density matrix evolve separately.
Following usual nomenclature, we refer to the diagonal matrix elements as \emph{populations} and to the off-diagonal as \emph{coherences}, although we emphasize that in the presence of degeneracy the distinction between the two is a matter of convention (a basis choice).

Because of the superselection rule~\eqref{eq:superselection_rule}, there are only two non-trivial coherences in the density matrix, one per parity sector:
\begin{subequations}\label{eq:coherences-definitions}
\begin{align}
c_\textrm{e} &= \bra{++}\rho\ket{--}\,,\\
c_\textrm{o} &= \bra{-+}\rho\ket{+-}\,.
\end{align}
\end{subequations}
In the even parity sector, we find 
\begin{equation}\label{eq:even_parity_coherence}
\frac{dc_\textrm{e}}{dt} = - \Gamma_\textrm{e}\,c_\textrm{e}
\end{equation}
with
\begin{equation}
\Gamma_\textrm{e} = \frac{\gamma(-U)}{2} = \Gamma_0\, (1+e^{\beta U})^{-1}.
\end{equation}
In the last equality, we have invoked Eq.~\eqref{eq:gamma_simple}. In the odd parity sector, instead, we find
\begin{equation}\label{eq:odd_parity_coherence}
\frac{dc_\textrm{o}}{dt} = - \Gamma_\textrm{o}\,c_\textrm{o}
\end{equation}
with
\begin{equation}
\Gamma_\textrm{o} = \frac{\gamma(U)}{2} = \Gamma_0\,(1+e^{-\beta U})^{-1}.
\end{equation}
We recover the crucial result: in the protected regime $\beta U\gg 1$, $\Gamma_\textrm{e}\approx \Gamma_0 e^{-\beta U}$ is exponentially suppressed compared to $\Gamma_0$. On the other hand, $\Gamma_\textrm{o}$ always remains $\approx\Gamma_0$.
The key factors behind this result are the presence of an energy gap in the spectrum and the detailed balance obeyed by the bath correlation function in thermal equilibrium.

The populations, defined as
\begin{equation}
p_{\sigma\tau} = \tr(\Pi_{\sigma\tau} \rho)=\bra{\sigma\tau}\rho\ket{\sigma\tau}\,,
\end{equation}
satisfy the following system of rate equations:
\begin{equation}
\label{eq:rate_equations_degenerate}
\frac{d}{dt}\begin{bmatrix}
p_{-+} \\ p_{+-} \\ p_{++} \\ p_{--}  
\end{bmatrix} = \frac{1}{2}
\begin{bmatrix}
-2\Gamma_\textrm{o} & 0 & \Gamma_\textrm{e} & \Gamma_\textrm{e} \\
0 & -2\Gamma_\textrm{o} & \Gamma_\textrm{e} & \Gamma_\textrm{e} \\
\Gamma_\textrm{o} & \Gamma_\textrm{o} & -2\Gamma_\textrm{e} & 0 \\
\Gamma_\textrm{o} & \Gamma_\textrm{o} & 0 & -2\Gamma_\textrm{e}
\end{bmatrix}
\begin{bmatrix}
p_{-+} \\ p_{+-} \\ p_{++} \\ p_{--}    
\end{bmatrix}\,.
\end{equation}
The steady state -- the unique zero mode of the rate equations -- is a thermal average of the even and odd sectors,
\begin{equation}\label{eq:rho_infty_degenerate}
\rho_\infty = \frac{(\Pi_{--}+\Pi_{++})+e^{-\beta U}(\Pi_{-+}+\Pi_{+-})}{2+2e^{-\beta U}}\,.
\end{equation}
The approach to the steady state is controlled by the non-zero eigenvalues of the rate equation matrix, which are $\Gamma_\textrm{o}$, $\Gamma_\textrm{e}$ and $\Gamma_\textrm{o}+\Gamma_\textrm{e}=\Gamma_0$.

Rather than presenting an explicit solution for $\rho(t)$, it more instructive to compute directly the time evolution of parity observables, which can be gathered from the above results. The expected value of the total parity $P$ obeys
\begin{equation}
\frac{d}{dt}\av{P}=\Gamma_\textrm{o}(1-\av{P})-\Gamma_\textrm{e}(1+\av{P})\,.
\end{equation}
The solution is
\begin{equation}
\av{P(t)} = e^{-\Gamma_0 t} \av{P(0)} + (1- e^{-\Gamma_0 t}) P_\infty
\end{equation}
with
\begin{equation}
P_\infty = \frac{1-e^{-\beta U}}{1+e^{-\beta U}}\,.
\end{equation}
We see that in the protected regime $\beta U \gg 1$ the total amount of leakage out of the qubit subspace is suppressed exponentially: $1-P_\infty\approx e^{-\beta U}$. This exponentially small amount of leakage occurs at a rate $\Gamma_0$.

The time evolution of the expectation values of the parity operators $P_{ij}=-2ia_ia_j$ is conditioned on the initial total parity $P$:
\begin{equation}
\av{P_{ij}(t)} = \begin{cases}
\av{P_{ij}(0)} \,e^{-\Gamma_\textrm{e} t} & \textrm{if}\quad P(0)=+1\,,\\
\\
\av{P_{ij}(0)} \,e^{-\Gamma_\textrm{o} t} & \textrm{if}\quad P(0)=-1\,.
\end{cases}
\end{equation}
These relations hold for all values of $i$ and $j$: since the qubit is encoded in two degenerate states, there is no distinction between relaxation and dephasing, and the decoherence of all qubit operators is set by the protected timescale $\Gamma_\textrm{e}$.

\subsection{Near degeneracy}
\label{sec:degenerate-near}

How do these results change in the presence of small couplings $A_{ij}$, which split the parity degeneracy of the energy spectrum?
We argue that there is a regime of weak overlaps, roughly speaking $\epsilon \ll \Gamma_0$, where results do not change qualitatively from the ideal case just presented.

First, if the energies $\epsilon_1$ and $\epsilon_2$ resulting from the tunneling between Majorana zero modes are small compared to $U$, then it is possible to approximate
\begin{equation}
\Gamma(U\pm \epsilon_1)\approx \Gamma(U \pm \epsilon_2) \approx \Gamma(U)\,.
\end{equation}
and similarly for $\Gamma(-U)$. In other words, the energy splittings do not affect considerably the transition strengths entering the dissipator of the Bloch-Redfield equation.

Second, the transition frequency differences $\delta_{ss'}$ listed in Eq.~\eqref{eq:transition_diffs} are all of the order of $\epsilon_1$ or $\epsilon_2$: non-zero but small in the scenario under consideration.
Thus, the resulting oscillations in the dissipator of the Bloch-Redfield are slow. The corresponding phase factors can be put to one, provided that $\abs{\delta_{ss'} \tau_D} \ll 1$, which means $\abs{\epsilon_1\pm \epsilon_2}\ll \Gamma_0$.

Under these conditions, the dissipator of the Bloch-Redfield equation remains the same as in the ideal case of Eq.~\eqref{eq:BR_ideal}.
Unlike the ideal case, however, the unitary dynamics of the tetron is not trivial, and coherent precession of the parity operators occurs with frequencies $\omega_\pm = \epsilon_1\pm \epsilon_2$, as discussed in Sec.~\ref{sec:unitary_dynamics}.

We note that since $\Gamma_\textrm{e} \ll \Gamma_0$ if $\beta U\gg 1$, the above-mentioned conditions are compatible with the condition $\Gamma_\textrm{e}\lesssim \epsilon$.
Thus, because the even parity sector decoheres very slowly, coherent oscillations may still be measurable in the time domain even when $\epsilon_1$ and $\epsilon_2$ may not be resolved in e.g. tunneling spectroscopy. 

\subsection{The zero-energy extended state scenario}
\label{sec:results_andreev}
We now consider a situation in which the Majorana zero  modes
$a_1$ and $a_2$ form an extended Andreev bound state at zero energy which can couple to two leads, while $a_3$ and $a_4$ remain well localized and coupled only to lead 3 and 4 respectively.
This scenario is interesting because extended Andreev bound states very close to zero energy can be tricky to distinguish from well-separated Majorana zero modes, as discussed for instance in Ref.~\cite{mishmash2020}.
Other than this modification to the system-bath couplings, we keep neglecting all direct couplings among the
Majorana modes, and so we remain in the degenerate regime.

In Appendix~\ref{app:correlation_functions} we show that, under some simplifying assumptions, this scenario can be modeled via the correlation function
\begin{equation}
\gamma_{ij}(\omega) = \gamma(\omega)\,(\delta_{ij} + \lambda\, \delta_{i1}\delta_{j2}+ \lambda\,\delta_{i2}\delta_{j1})\,. 
\end{equation}
Here, $\lambda$ is a dimensionless parameter taking values in the interval $[-1, 1]$.
It controls the relative strength of the tunneling rates between $a_1$ and $a_2$ and the leads.
When $\lambda=0$, $a_1$ is coupled only to lead 1, while when $\abs{\lambda}=1$, $a_1$ is coupled equally strongly to leads 1 and 2.
The same applies to the tunneling rates between $a_2$ and the leads.
We note that this parametrization preserves the positivity condition of $\gamma_{ij}(\omega)$.

The simplifying assumptions behind this form of the system-bath couplings are the following. First, as in the previous calculations in this section, we have assumed identical tunneling rates between $a_i$ and lead $i$. Second, we have assumed that the density of states in the leads is particle-hole symmetric, $\nu(\epsilon)=\nu(-\epsilon)$. This, in particular, allows $\lambda$ to be real. 

The presence of an extended state gives rise to a non-zero value of $\lambda$ and thus to a finite
cross-correlation $\gamma_{12}(\omega)$.
The master equation in the degenerate regime must be supplemented by a new term;
\begin{equation}
\frac{d\rho}{dt} = \mathcal{D}_0[\rho] + \lambda\,\mathcal{D}_1[\rho]
\end{equation}
with
\begin{align}\nonumber
\mathcal{D}_1[\rho] &= \frac{\gamma(U)}{4}\,\left[a_2\rho_\textrm{o}\,a_1 + a_1\rho_\textrm{o}\,a_2\right]\\
&+\frac{\gamma(-U)}{4}\,\left[a_2\rho_\textrm{e}\,a_1 + a_1\rho_\textrm{e}\, a_2\right]\,.
\end{align}
The addition of $\mathcal{D}_1[\rho]$ does not modify the rate equations for the populations, Eq.~\eqref{eq:rate_equations_degenerate}.
However, it affects the time evolution of the coherences $c_\textrm{e}$ and $c_\textrm{o}$ defined in Eq.~\eqref{eq:coherences-definitions}.
Namely, Eqs.~\eqref{eq:even_parity_coherence} and~\eqref{eq:odd_parity_coherence} must be supplemented with the following \emph{coupled} system of equations
\begin{equation}
\frac{d}{dt}\begin{bmatrix} c_\textrm{o} \\ c_\textrm{e}\end{bmatrix} = \frac{1}{2}
\begin{bmatrix}
-2\Gamma_\textrm{o} & i\lambda \Gamma_\textrm{e} \\
-i\lambda\Gamma_\textrm{o} & -2\Gamma_\textrm{e}
\end{bmatrix}\begin{bmatrix} c_\textrm{o} \\ c_\textrm{e}\end{bmatrix}\,.
\end{equation}
We see that a finite $\lambda$ induces a transfer of coherence between the two parity sectors.
Indeed, $c_\textrm{o}$ and $c_\textrm{e}$ do not evolve separately under the decay rates $\Gamma_\textrm{o}$ and $\Gamma_\textrm{e}$, but jointly under the new decay rates (eigenvalues of the $2\times 2$ matrix above)
\begin{equation}
\Gamma_{\pm}=\tfrac{1}{2}\Gamma_0\,(1\pm \kappa)
\end{equation}
with
\begin{equation}
\kappa^2 = \left(\frac{\Gamma_\textrm{o}-\Gamma_\textrm{e}}{\Gamma_0}\right)^2 + \lambda^2\,\frac{\Gamma_\textrm{e}\Gamma_\textrm{o}}{\Gamma_0^2}\,.
\end{equation}

Let us imagine that we initialize the system in the even parity sector, so that $c_\textrm{o}(0)=0$. Then, the time evolution of the coherences is
\begin{align}
c_\textrm{o}(t) &= -i c_\textrm{e}(0)\,\frac{\lambda}{\kappa}\frac{\Gamma_\textrm{e}}{\Gamma_0}\,\exp\left(-\tfrac{1}{2}\Gamma_0 t\right)\,\sinh\left(\tfrac{1}{2}\kappa\Gamma_0 t\right)\,,\\\nonumber
c_\textrm{e}(t) &= c_\textrm{e}(0)\,\left[\left(\frac{\kappa\Gamma_0 + \Gamma_\textrm{e}-\Gamma_\textrm{o}}{2\kappa\Gamma_0}\right) \exp\left(-\Gamma_+t\right)\right.\\
&\qquad\quad+\left.\left(\frac{\kappa\Gamma_0 - \Gamma_\textrm{e}+\Gamma_\textrm{o}}{2\kappa\Gamma_0}\right)\exp\left(-\Gamma_-t\right)\right]\,.
\end{align}
We see that a finite $\lambda$ turns the simple exponential decay of the coherence in Sec.~\ref{sec:degenerate_ideal} into a bi-exponential decay, because it couples two decoherence channels which were previously independent.

Note that, at finite $\lambda$, one has $\Gamma_+ > \Gamma_\textrm{o}$ and $\Gamma_{-} < \Gamma_\textrm{e}$. In other words, increasing $\abs{\lambda}$ from zero, $\Gamma_+$ tends to increase from $\Gamma_\textrm{o}$, while $\Gamma_-$ decreases from $\Gamma_\textrm{e}$, reaching $\Gamma_-=\tfrac{1}{2}\Gamma_\textrm{e}$ at $\lambda=1$.
Due to the decrease in $\Gamma_{-}$, we can say that the cross-correlation slows down the decoherence.
In fact, in a scenario where $a_3$ and $a_4$ are also extended zero modes, one can tune the rate $\Gamma_-$ to be arbitrarily close to zero~\cite{bhattacharyya2026}.

How do these results on populations and coherences translate into the evolution of parity observables? Because the rate equations for the populations are independent of $\lambda$, the time dependence of the total parity $P$ and the ``longitudinal'' parities $P_{12}$ and $P_{34}$ is unaffected with respect to the results of Sec.~\ref{sec:degenerate_ideal}. On other hand, the ``transverse'' parity observables $P_{13}, P_{14}, P_{23}, P_{24}$ inherit the bi-exponential decay of the coherences.

\section{The secular regime}
\label{sec:secular}

We now analyze the case in which the energies $\epsilon_1$ and $\epsilon_2$ become larger than the system-bath coupling strength:
\begin{equation}
\epsilon \gg \Gamma_0\,.
\end{equation}
In this regime, the Bohr transition frequencies of the tetron can be resolved by the bath.
It is then appropriate to invoke the conventional \emph{secular approximation}, in which one neglects all terms in the Bloch-Redfield equation except those with $\delta_{ss'}=0$.
The idea is that for all terms with non-zero $\delta_{ss'}$, the inequality
\begin{equation}
\abs{\delta_{ss'}\tau_D} \sim \epsilon\, \Gamma^{-1}_0 \gg 1
\end{equation}
holds, and so the corresponding term averages out to zero over an evolution time of order $\tau_D$.

To be concrete and keep the calculations transparent and analytically tractable, in this section we consider a simpler version of the system Hamiltonian $H_A$:
\begin{equation}\label{eq:HA_simple}
H_A = i\epsilon (a_1a_2+a_3a_4) + 2U a_1 a_2 a_3 a_4\,.
\end{equation}
In this version, the computational Majorana zero modes coincide with the eigenmodes, and furthermore we have $\epsilon_1=\epsilon_2=\epsilon$.
This choice introduces a finite energy splitting in the even-parity qubit subspace, the essential feature that we would like to capture in our results.
On the other hand, the odd-parity subspace remains doubly degenerate, which eliminates unessential complications in the calculations.
This scenario also captures a situation -- relevant for current experimental designs -- in which the tetron is realized from two distinct topological nanowire segments, the first hosting zero modes $a_1$ and $a_2$ and the second hosting $a_3$ and $a_4$, with inter-wire overlaps all but negligible.

For this simple version of the system Hamiltonian, the computational Majorana zero modes $a_i$ coincide with the eigenmodes $\tilde{a}_i$, and the transition frequency differences split in two groups: half of them are equal to $2\epsilon$, while the rest is equal to zero.
The Bloch-Redfield equation can then be simplified considerably: with some algebra and patience - displayed in App.~\ref{app:secular} -- invoking the secular approximation then leads to the master equation: 
\begin{equation}\label{eq:secular_general_me}
\frac{d\rho}{dt}= \mathcal{D}_\textrm{sec}[\rho].
\end{equation}
The dissipator $\mathcal{D}_\textrm{sec}$, for disconnected and identical reservoirs such that
$\gamma_{ij}(\omega)=\delta_{ij}\gamma(\omega)$, is
\begin{widetext}
\begin{align}\nonumber
\mathcal{D}_\textrm{sec}[\rho]=&-\frac{1}{4}\gamma(U-\epsilon)\,\left[\rho_\textrm{o} -\sum_i \Pi_{--}\,a_i\,\rho\,a_i\,\Pi_{--}\right]-\frac{1}{4}\gamma(\epsilon-U)\,\left[\{\Pi_{--},\rho\}-\sum_i a_i\,\Pi_{--}\,\rho\,\Pi_{--}\,a_i\right]\\
&-\frac{1}{4}\gamma(U+\epsilon)\,\left[\rho_\textrm{o} - \sum_i \Pi_{++}\,a_i\,\rho\,a_i\,\Pi_{++}\right]-\frac{1}{4}\gamma(-\epsilon-U)\,\left[\{\Pi_{++},\rho\}-\sum_i a_i\,\Pi_{++}\,\rho\,\Pi_{++}\,a_i\right]\,.
\label{eq:dissipator_secular}
\end{align}
\end{widetext}

Unlike in the degenerate case discussed in the previous Section~\ref{sec:degenerate}, in this regime the distinction between population and coherences is physical: the common eigenstates of $P_{12}$ and $P_{34}$ provide a preferred basis as eigenstates of $H_A$. The advantage of having chosen this basis already to discuss the degenerate case is that the formulae below can be compared directly to the analogous one provided in Sec.~\ref{sec:degenerate}.

In fact the coherences $c_\textrm{e}$ and $c_\textrm{o}$ obey the same evolution as before, see respectively Eq.~\eqref{eq:even_parity_coherence} and Eq.~\eqref{eq:odd_parity_coherence}, except that the decoherence rates $\Gamma_\textrm{e}$ and $\Gamma_\textrm{o}$ must be updated. In the even parity sector, we find
\begin{equation}
\Gamma_\textrm{e} = \frac{1}{4}\left[\gamma(\epsilon-U)+\gamma(-\epsilon-U)\right]\,.
\end{equation}
Using Eq.~\eqref{eq:gamma_simple}, this expression reduces to
\begin{equation}\label{eq:decoherence_even_secular}
\Gamma_\textrm{e} = \frac{\Gamma_0}{2}\,\left[\frac{1}{1+e^{\beta (U-\epsilon)}}+\frac{1}{1+e^{\beta (U+\epsilon)}}\right]
\end{equation}
If $U>\epsilon$ and additionally $\beta(U-\epsilon)\gg 1$, the rate is well approximated by
\begin{equation}
\Gamma_\textrm{e} \approx \Gamma_0\,e^{-\beta U}\,\cosh(\beta\epsilon)\,.
\end{equation}
As we can see, in this regime the suppression of decoherence due to the interaction $U$ is softened by the energy splitting $\epsilon$. As $\epsilon$ increases, the topological protection is compromised. The exponential suppression of the decoherence rate disappears, even at low temperatures, once $\epsilon$ becomes comparable to or larger than $U$.

In the odd-parity sector, instead, we find
\begin{equation}
\Gamma_\textrm{o} = \frac{1}{4}\left[\gamma(U-\epsilon)+\gamma(U+\epsilon)\right]\,.
\end{equation}
Using again Eq.~\eqref{eq:gamma_simple}, this expression becomes
\begin{equation}
\Gamma_\textrm{o} = \frac{\Gamma_0}{2}\,\left[\frac{1}{1+e^{-\beta (U-\epsilon)}}+\frac{1}{1+e^{-\beta (U+\epsilon)}}\right]\,.
\end{equation}
which is never exponentially smaller than $\Gamma_0$.

Conditioned on the initial total parity being even or odd, the decoherence rates $\Gamma_\textrm{e}$ and $\Gamma_\textrm{o}$ also control the time evolution of the ``transverse'' parity observables $P_{13}$, $P_{14}$, $P_{23}$ and $P_{24}$ -- those not associated with population of the energy levels. 

The time evolution of the populations $p_{\sigma\tau}$ is more complex than in the degenerate case. Starting from the master equation~\eqref{eq:secular_general_me}, we find the following system of rate equations, which replaces Eq.~\eqref{eq:rate_equations_degenerate}:
\begin{widetext}
\begin{equation}\label{eq:rate_equations_secular}
\frac{d}{dt}\begin{bmatrix}
p_{--} \\ p_{+-} \\ p_{-+} \\ p_{++}  
\end{bmatrix} = \frac{1}{4}\,
\begin{bmatrix}
-2\gamma(\epsilon-U) & \gamma(U-\epsilon) & \gamma(U-\epsilon) & 0 \\
\gamma(\epsilon-U) & -\gamma(U-\epsilon)-\gamma(U+\epsilon) & 0 & \gamma(-\epsilon-U) \\
\gamma(\epsilon-U) & 0 & -\gamma(U-\epsilon)-\gamma(U+\epsilon) & \gamma(-\epsilon-U) \\
0 & \gamma(U+\epsilon) & \gamma(U+\epsilon) & -2\gamma(-\epsilon-U)
\end{bmatrix}
\begin{bmatrix}
p_{--} \\ p_{+-} \\ p_{-+} \\ p_{++}    
\end{bmatrix}\,.
\end{equation}
\end{widetext}
The steady state is now a Gibbs thermal state that takes into account the energy splitting $\epsilon$,
\begin{equation}\label{eq:rho_infty_secular}
\rho_\infty = \frac{e^{-\beta\epsilon}\Pi_{--}+e^{-\beta U}(\Pi_{-+}+\Pi_{+-}) + e^{\beta \epsilon}\Pi_{++}}{2\cosh(\beta \epsilon)+2e^{-\beta U}}\,.
\end{equation}
From the populations, we can obtain also the time evolution of the total parity $P$ as well as the ``longitudinal'' parities $P_{12}$ and $P_{34}$, since these parity observables are just linear combinations of projectors:
\begin{subequations}
\begin{align}
P &= \Pi_{++} + \Pi_{--} - \Pi_{-+} - \Pi_{+-} \,,\\
P_{12} &= \Pi_{+-} + \Pi_{++} - \Pi_{--} - \Pi_{-+}\,,\\
P_{34} &= \Pi_{-+} + \Pi_{++} - \Pi_{--} - \Pi_{+-}\,.
\end{align}
\end{subequations}
In particular, from Eq.~\eqref{eq:rho_infty_secular} we can easily obtain their steady-state expectation values,
\begin{equation}
P_\infty = \frac{\cosh(\beta\epsilon)-e^{-\beta U}}{\cosh(\beta\epsilon)+e^{-\beta U}}\,,
\end{equation}
and
\begin{equation}
P_{12}^\infty = P_{34}^\infty = \frac{\sinh(\beta\epsilon)}{\cosh(\beta\epsilon)+e^{-\beta U}}
\end{equation}
The approach to these steady state values is controlled by the non-zero eigenvalues of the matrix in Eq.~\eqref{eq:rate_equations_secular}. Using Eq.~\eqref{eq:gamma_simple}, these are given -- analogously to the degenerate case -- by $\Gamma_\textrm{o}$, $\Gamma_\textrm{e}$ and their sum $\Gamma_\textrm{o}+\Gamma_\textrm{e}=\Gamma_0$.

The time evolution of the ``longitudinal'' coherences $P_{12}$ and $P_{34}$, though quite convoluted to write for arbitrary initial states, becomes simple if we assume that the initial state is one of the two qubit states $\ket{++}$ or $\ket{--}$. Then, we have
\begin{equation}
\av{P_{12}(t)} = \av{P_{34}(t)} = \pm e^{-\Gamma_\textrm{e}t} + (1-e^{-\Gamma_\textrm{e}t}) P_{12}^\infty.
\end{equation}
and so we see that the relaxation to the steady state occurs over the same timescale that dictates the dephasing of the qubit.

\subsection{Relation to the universal Lindblad equation}
\label{sec:ULE}

It is quite immediate to notice that the dissipator in the secular regime, Eq.~\eqref{eq:dissipator_secular}, is smoothly connected to that in the degenerate regime, Eq.~\eqref{eq:dissipator_degenerate}, as one sends $\epsilon\to 0$.
Naturally, this is also true of all the decoherence rates derived from the secular dissipator.
One is therefore tempted to extend the regime validity of the formulae derived in the secular regime beyond the one in which they were originally derived, namely $\epsilon \gg \Gamma_0$.

For this purpose, we compared the secular master equation derived above with the universal Lindblad equation (ULE)~\cite{nathan2020}.
The ULE is a positive and Markovian master equation: it is derived under the Born-Markov approximation $\tau_D\gg\tau_B$, but without making any type of approximation (such as the secular one $\epsilon\gg\Gamma_0$) which requires relating the spectral properties of the system with the strength of the system-bath coupling. In the context of Majorana zero modes, the ULE has been already applied to the study of tetron devices in Refs.~\cite{munk2020,boutin2025}.

For disconnected and identical reservoirs,
$\gamma_{ij}(\omega)=\delta_{ij}\gamma(\omega)$, the ULE dissipator
takes the form
\begin{equation}
    \mathcal{D}_{\rm ULE}[\rho]
    =
    -\frac{1}{4}\sum_{i=1}^4
    \big(
        K_i\rho K_i^\dagger
        -\frac{1}{2}\{K_i^\dagger K_i,\rho\}
    \big),
    \label{eq:ULE_dissipator}
\end{equation}
with jump operators
\begin{equation}
    K_i =
    \sum_{s=1}^4
    \left[
        \sqrt{\gamma(\omega_s)}\,A_{i,s}
        +
        \sqrt{\gamma(-\omega_s)}\,A_{i,s}^{\dagger}
    \right].
    \label{eq:ULE_jump}
\end{equation}
In contrast to the secular master equation, a single ULE jump operator is composed of transitions at different Bohr frequencies and therefore generates interference terms between them in general.
For the model considered in this Section,
Eq.~\eqref{eq:ULE_jump} becomes
\begin{align}
    K_i={}&
    \sqrt{\gamma(U-\epsilon)}\,\Pi_{--}a_i
    +\sqrt{\gamma(\epsilon-U)}\,a_i\Pi_{--}
    \nonumber\\
    &+
    \sqrt{\gamma(U+\epsilon)}\,\Pi_{++}a_i
    +\sqrt{\gamma(-U-\epsilon)}\,a_i\Pi_{++}\,. .
    \label{eq:ULE_jump_special}
\end{align}
As shown in Appendix~\ref{app:ULE}, all interference terms in
$K_i^\dagger K_i$ vanish identically. In the recycling term
$K_i\rho K_i^\dagger$, terms connecting different total-parity
blocks vanish for density matrices satisfying $[\rho,P]=0$.
The remaining interference terms, which couple the two distinct
Bohr frequencies $U-\epsilon$ and $U+\epsilon$, cancel after
summing over the two Majoranas within each canonical pair.
Consequently,
\begin{equation}
    \mathcal{D}_{\rm ULE}[\rho]
    =
    \mathcal{D}_{\rm sec}[\rho]\,.
    \label{eq:ULE_sec_equivalence}
\end{equation}
This equivalence shows that the regime of validity of the results obtained from the secular dissipator can be extended all the way to $\epsilon=0$, and thus that they are smoothly connected with the results obtained in the degenerate regime, Sec.~\ref{sec:degenerate}.

\section{Conclusions}
\label{sec:conclusions}

We have studied the dissipative dynamics of a Majorana tetron qubit under the presence of (extrinsic) quasiparticle poisoning, within the Born-Markov approximation.
Our results are summarized by a formula for the decoherence rate of a Majorana qubit, Eq.~\eqref{eq:decoherence_even_secular}:
\begin{equation}
\Gamma_\textrm{e} = \frac{\Gamma_0}{2}\,\left[\frac{1}{1+e^{\beta (U-\epsilon)}}+\frac{1}{1+e^{\beta (U+\epsilon)}}\right]
\end{equation}
The formula has a humble explanation: it is the absorption rate of a quasiparticle from the leads, averaged over the two even-parity qubit states.
It illustrates the crossover outside of the topologically protected regime $\epsilon \ll \Gamma_0 \ll T \ll U$ as either the temperature or the energy splitting $\epsilon$ are increased.
Considering the argument just made in Sec.~\ref{sec:ULE}, its regime of validity is quite broad: the formula applies to arbitrary ratios $\epsilon/\Gamma_0$, $\epsilon/U$ and $U/T$, provided that $T \gg \Gamma_0$.
The decoherence rate above applies to all qubit observables (i.e., bilinear parity operators) provided the tetron is initialized in an even parity state.
In Sec.~\ref{sec:degenerate} and Sec.~\ref{sec:secular}, we collected more analytical results on the rate equations, steady state and leakage rate governing the time evolution of the qubit.

If some of the Majorana zero modes have extended wave functions that couple to more than one lead, two or more relaxation channels become coupled and the decay of the decoherence transitions from a simple exponential to a multi-exponential (see Sec.~\ref{sec:results_andreev}). This may help distinguish well-localized Majorana zero modes from zero-energy Andreev states which are extended throughout the entire device. It is amusing that a signature of the latter is a \emph{slowdown} of decoherence due to quasiparticle poisoning; although we have to add that an extended state is sensitive to other local sources of noise -- not included in our analysis -- to which well-separated Majorana zero modes are not.

In our derivations, we have focused on a few scenarios that isolated qualitatively important factors -- such as the finite energy splitting between qubit states and the presence of cross-correlations between leads.
In doing so, we have allowed ourselves to make simplifying assumptions -- such as taking individual tunneling rates for all the leads to be identical, or the energy splitting of odd-parity states to be zero -- in order to arrive at simple analytical results that can guide understanding of the general case. The latter is probably amenable only to numerical analysis, which we leave for future work.

Our theory does not apply to \emph{intrinsic} quasiparticle poisoning~\cite{karzig2021,alase2025} related to the excitation of above-gap quasiparticles inside the topological superconducting island, nor to \emph{extrinsic} quasiparticle poisoning due to a \emph{non-equilibrium} population of quasiparticles in the external reservoirs.
In fact, the detailed balance of Eq.~\eqref{eq:detailedbalance} is the essential assumption which underpins the above derivation of exponentially protected decoherence rates.

The study of the open-system dynamics of tetron devices in the presence of a non-equilibrium distribution of quasiparticles represents an important theoretical problem to be tackled, for which the present work provides a baseline.
Another interesting direction is the study of the error induced by quasiparticle poisoning on braiding operations, extending previous work based on parity-preserving noise sources~\cite{pedrocchi2015,lai2020,sahu2026}.

\subsection*{Data availability}

No data was generated or analyzed during this work.

\subsection*{AI statement}

We have used Gemini to generate the tikz source code for the figures in the main text, starting from our own hand-drawn sketches.

\begin{acknowledgments}
We thank Marco Grilli for nice and useful discussions on the project. S.B. acknowledges financial support from the PNRR MUR project PE0000023-NQSTI, and specifically the project ‘Topological Phases of Matter, Superconductivity, and Heterostructures’ Partenariato Esteso 4-Spoke 5 (n. PE4221852A63A88D).
\end{acknowledgments}

\appendix

\section{Derivation of the Bloch-Redfield equation}
\label{sec:app-B}

The derivation of the Bloch-Redfield of Eq.~\ref{eq:Bloch-Redfield} starting from the model Hamiltonian of Sec.~\ref{sec:model} broadly follows the standard steps described e.g. in Ref.~\cite{breuer2002}. We have already described it in detail for the case $U=0$ in our previous work~\cite{bhattacharyya2026}, and the first part of that derivation remains valid at finite $U$. Using the interaction picture with respect to $H_A+H_B$, truncating the von Neumann equation at second order in the system-bath coupling, taking the trace over the bath and employing the Born-Markov approximation, one arrives at:
\begin{align}\label{eq:born-markov_me}\nonumber
\dot \rho = -\frac{1}{4}\sum_{i,j=1}^{4}&\bigg\{\int_0^\infty dt'\,C_{ij}(t')\,[a_i(t),a_j(t-t')\,\rho(t)] \\ +& \int_0^\infty dt'\,C_{ji}(-t')\,[\rho(t)\,a_j(t-t'),a_i(t)]\bigg\}.
\end{align}
At this point we make use of the jump operators $A_{i,s}$ introduced in Eq.~\eqref{eq:jump_operators}. They decompose a Majorana operator into a sum corresponding to individual transitions between eigenstates:
\begin{equation}
a_i = \sum_{s} (A_{i,s} + A^\dagger_{i,s})\,.
\end{equation}
We recall that $A_{i,s}$ are attached to transitions from odd-parity initial states to even-parity final states, and vice-versa $A^\dagger_{i,s}$ are attached to even-to-odd transitions.
These operators have the nice property that
\begin{equation}
[H_{A}, A_{i,s}] = -\omega_s A_{i,s}\, .
\end{equation}
This enables one to write $a_{i}(t)$ as
\begin{equation}
a_{i}(t) = \sum_s \left[\exp(-i\omega_s t) A_{i,s} + \textrm{h.c.}\right]\,.
\end{equation}
The next step is to put back this expression for $a_{i}$ into the master equation~\eqref{eq:born-markov_me}.
When doing so, we can make two considerations to reduce the number of terms that remain in the result.
First, when there are two successive $A$ operators multiplying $\rho$, they must correspond to opposite order of parity flips - so one $A$ and one $A^\dagger$.
Second, in the terms in which $\rho$ is sandwiched between two $A$ operators, the latter again should be of the opposite type if the initial $\rho$ has no parity coherences, i.e. if $[\rho, P]=0$ as discussed in the main text.

These selection rules considerably simplify the frequency mixing structure of the resulting master equation.
Taking them into account, we obtain the identities
\begin{widetext}
\begin{align}\nonumber
[a_i(t),a_j(t-t')\rho] &= \sum_{s,s'} \left[e^{-i(\omega_s-\omega_{s'})t}e^{ i\omega_s t'} (A^\dagger_{i,s'}A_{j,s}\rho - A_{j,s} \rho A^\dagger_{i,s'}) + \textrm{h.c.}\right],\\ 
[\rho a_j(t-t'),a_i(t)] &= \sum_{s,s'} \left[e^{i(\omega_s-\omega_{s'})t}\,e^{-i\omega_s t'} (\rho A_{j,s}A^\dagger_{i,s'} - A^\dagger_{i,s'} \rho A_{j,s})+\textrm{h.c.}\right]\,.
\end{align}
Plugging these back into Eq.~\eqref{eq:born-markov_me}, introducing the frequency differences $\delta_{ss'}=\omega_{s'}-\omega_s$, performing the integral over $t'$, and defining
\begin{equation}
\Gamma_{ij}(\omega) = \int_{0}^{\infty} dt \exp(i\omega t) C_{ij}(t)\,,
\end{equation}
one finally obtains:
\begin{align}\label{eq:Bloch-Redfield-ij}
\dot \rho &= -\frac{1}{4} \sum_{i,j}\sum_{s,s'} \left[e^{i\delta_{ss'}t}\,\Gamma_{ij}(\omega_s) (A^{\dagger}_{i,s'}A_{j,s}\rho - A_{j,s}\, \rho\, A^{\dagger}_{i,s'}) + e^{-i\delta_{ss'}t}\,\Gamma_{ji}(-\omega_s) (A_{i,s'}A^{\dagger}_{j,s}\rho - A^{\dagger}_{j,s}\, \rho\, A_{i,s'}) + \textrm{h.c.}\right]\,,
\end{align}
\end{widetext}
of which we present in the main text the simplified version with $\Gamma_{ij}(\omega)=\delta_{ij}\Gamma(\omega)$. The more general decomposition of $\Gamma_{ij}(\omega)$ is $\Gamma_{ij}(\omega)=\tfrac{1}{2}\gamma_{ij}(\omega)+iS_{ij}(\omega)$ with
\begin{align}
\gamma_{ij}(\omega) &= \Gamma_{ij}(\omega) + \Gamma_{ji}^{*}(\omega)\,,\\
S_{ij}(\omega) &= [\Gamma_{ij}(\omega)-\Gamma_{ji}^{*}(\omega)]/2i\,. 
\end{align}
Using the above definitions and simplifying the master equation~\ref{eq:born-markov_me}, we finally obtain the equation of the Bloch-Redfield form (Eq.~\ref{eq:Bloch-Redfield}), quoted in Sec.~\ref{sec:bloch-redfield}.

\section{Bath correlation functions}
\label{app:correlation_functions}

To complete the appendix, we provide a more explicit expression for the bath correlation function. Starting from Eq.~\eqref{eq:Cij}, a standard equilibrium calculation yields
\begin{align}\label{eq:reservoir correlation}
C_{ij}(t) = \sum_l &\left[w_{il} w^*_{jl}\int_{-\infty}^\infty  \nu_l(\epsilon)\,e^{-i\epsilon t} f(-\epsilon)\,d\epsilon\right.\\\nonumber
&\quad\left.+w^*_{il}w_{jl} \int_{-\infty}^\infty \nu_l(\epsilon)\,e^{i\epsilon t} f(\epsilon)\,d\epsilon\right]\,,
\end{align}
where $f(\epsilon)$ is the Fermi-Dirac distribution and $\nu_i(\epsilon)$ is the local density of states in reservoir $i$ at the point contact position $\mathbf{r}_i$:
\begin{equation}
\nu_i(\epsilon) = \sum_k \abs{\phi_k(\mathbf{r}_i)}^2\,\delta(\epsilon-\zeta_k)\,.
\end{equation}
Here, the index $k$ runs over a complete set of single-particle states of the reservoir, with energies $\zeta_k$. For simplicity we will assume $\nu_i(\epsilon)\equiv \nu(\epsilon)$, the same for all four reservoirs.
Then, we can decompose the correlation function as
\begin{align}
C_{ij}(t) = z_{ij} \int_{-\infty}^\infty  \nu(\epsilon)\,e^{-i\epsilon t} f(-\epsilon)\,d\epsilon + z_{ij}^* \int_{-\infty}^\infty \nu(\epsilon)\,e^{i\epsilon t}
\end{align}
with
\begin{equation}
z_{ij} = \sum_l w_{il} w^*_{jl}\,,
\end{equation}
a complex number with the property $z_{ji}=z_{ij}^*$. By Fourier transform, we then obtain
\begin{align}\label{eq:gamma_ij_general}
\gamma_{ij}(\omega) =& 2\pi \left[z_{ij} \nu(\omega) +  z_{ij}^* \nu(-\omega)\right]\,f(-\omega)\,.
\end{align}
This function obeys the detailed balance condition
\begin{equation}
\gamma_{ji}(-\omega) = e^{-\beta\omega}\gamma_{ij}(\omega)\,.
\end{equation}
Finally, using the fact that $C_{ij}(-t)=C^{*}_{ji}(t)$, we also find
\begin{equation*}
S_{ij}(\omega) = -\frac{1}{2\pi}\,\textrm{p.v.}\int_{-\infty}^\infty \frac{\gamma_{ij}(\omega')}{\omega'-\omega}\,d\omega'\,,
\end{equation*}
a Kramers-Krönig relation.

\emph{Disconnected reservoirs.} In the simple case $w_{ij}=w\,\delta_{ij}$, often adopted in the main text, we have
\begin{equation}
z_{ij}=\abs{w}^2\delta_{ij}\,.
\end{equation}
In this case, $C_{ij}(t)=\delta_{ij} C(t)$ with
\begin{equation}\label{eq:C(t)}
C(t) = \abs{w}^2 \int_{-\infty}^\infty d\epsilon\,\nu(\epsilon)
 \left[ \cos(\epsilon t)  -
 i\tanh\left(\tfrac{1}{2}\beta\epsilon\right)
 \sin(\epsilon t)
 \right]\,,
\end{equation}
and
\begin{equation}
\gamma(\omega) = 2\pi \abs{w}^2\,\frac{\nu(\omega)+\nu(-\omega)}{1+e^{-\beta\omega}}\,.
\end{equation}

\emph{Correlation functions for Sec.~\ref{sec:results_andreev}.} The case of a zero-energy extended state case corresponds to the choice
\begin{equation}
w_{ij}=w_{i} \delta_{ij}+w_{12}\, \delta_{i1}\delta_{j2} + w_{21}\,\delta_{i2}\delta_{j1}\,.
\end{equation}
for some complex numbers $w_i, w_{12}, w_{21}$. In this case, we have
\begin{align}
z_{11} &= \abs{w_1}^2 + \abs{w_{12}}^2\,,\\
z_{12} &= (w_{1}w_{21}^*+w_{12}w_{2}^*) = z_{21}^*\,,\\
z_{22} &= \abs{w_2}^2 + \abs{w_{21}}^2\,,\\
z_{33} &= \abs{w_3}^2 \,,\\
z_{44} &= \abs{w_4}^2 \,.
\end{align}
The functions $\gamma_{ij}(\omega)$ can be obtained by replacement into Eq.~\eqref{eq:gamma_ij_general}. It can be shown that $\gamma_{ij}(\omega)$ is positive if
\begin{equation}
z_{11}z_{22}\geq \abs{z_{12}}^2\,,
\end{equation}
which is true for any choice of the parameters $w_i, w_{12}, w_{21}$. To simplify the mathematical treatment of this scenario, we then choose a symmetric situation in which $z_{11}=z_{22}=z_{33}=z_{44}=\abs{w}^2$ and $z_{12}= \lambda \abs{w}^2$. The positivity condition is satisfied if $\abs{\lambda}\leq 1$. With this choice, one has
\begin{equation}
\gamma_{12}(\omega) = 2\pi \abs{w}^2 [\lambda\,\nu(\omega) + \lambda^* \nu(-\omega)]\,f(-\omega)\,.
\end{equation}
If we further assume that $\nu(\omega) = \nu(-\omega)$, then this expression simplifies to
\begin{equation}
\gamma_{12}(\omega) = \lambda \gamma(\omega)
\end{equation}
with $\lambda$ now real, and $-1\leq \lambda \leq 1$. In this case, one finally gets
\begin{equation}
\gamma_{ij}(\omega) = \gamma(\omega)\,(\delta_{ij} + \lambda\, \delta_{i1}\delta_{j2}+ \lambda\,\delta_{i2}\delta_{j1})\,.
\end{equation}

\section{Example of a wide-band fermionic bath}
\label{app:wide-band}

In Sec.~\ref{sec:model}, we introduced the bath correlation time $\tau_B$
and the dissipative timescale $\tau_D$ without specifying a
microscopic model for the leads.
In this Appendix, using a simple toy model for the fermionic reservoir, we provide explicit estimates of these two timescales, illustrating the separation underlying the Born--Markov approximation used throughout the paper.

\begin{figure}[t!]
    \centering
    \includegraphics[width=\linewidth]{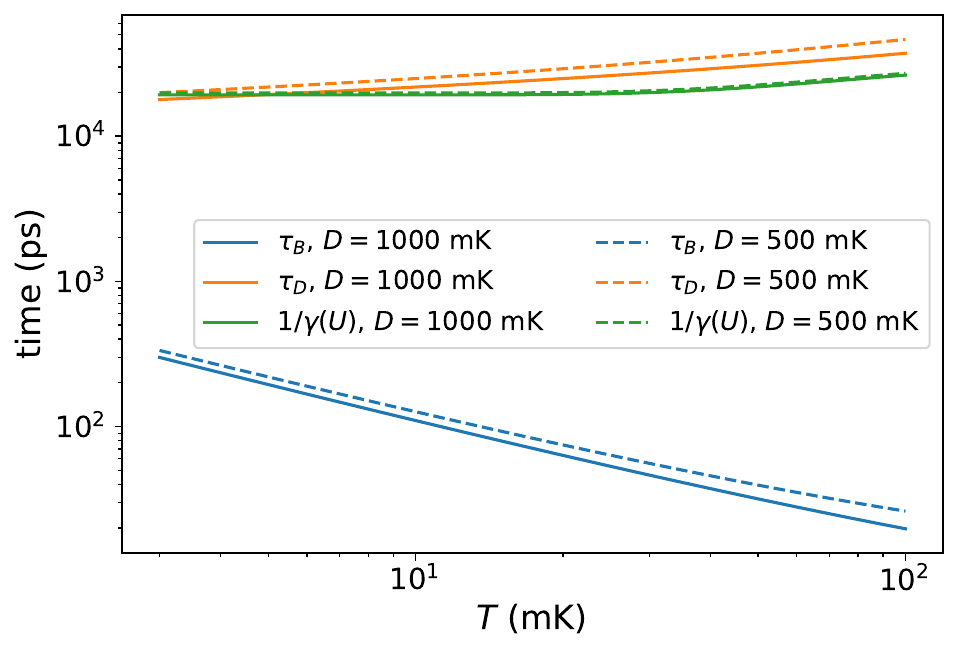}
    \caption{Plots of the numerically calculated $\tau_{B}$, $\tau_{D}$, and $\gamma^{-1}(U)$, for the wide-band model of the bath. The two sets of results (in solid and dashed lines) are for two choices of $D=1$K, and $D=0.5$K respectively. The value of $U=100mK$ for both curves. We observe a clear separation of scales $\tau_{B} \ll \tau_{D}$, and also the fact that $\gamma^{-1}(U) \sim \tau_{D}$ from the plots.}
    \label{fig:bath_timescales}
\end{figure}

To obtain explicit estimates, we apply the formulae derived in the previous Appendix to the case in which the density of states is given by a broad Lorentzian,
\begin{equation}
 \nu(\epsilon) =  \frac{\nu(0)\,D^2}{\epsilon^2+D^2}\,,
 \label{eq:app_JLorentzian}
\end{equation}
where $D$ is a soft bandwidth which provides a smooth realization of the wide-band limit. In particular, when
\begin{equation}
 D \gg U,\;\epsilon,\;T\,,
 \label{eq:app_wideband}
\end{equation}
the spectral density varies weakly over the energy window relevant
to the tetron dynamics. In the near-degenerate regime this allows, in
particular, the replacement
\begin{equation}
 \gamma(U\pm\epsilon_i)\simeq\gamma(U),
 \label{eq:app_gammaflat}
\end{equation}
discussed in Sec.~\ref{sec:degenerate-near}.

We next consider the two timescales $\tau_B$ and $\tau_D$ introduced in Sec.~II, Eqs.~\eqref{eq:tauB} and Eq.~\eqref{eq:tauD} respectively.
In the main text, we argued that $\tau_B^{-1}\approx k_BT$ whereas $\tau_D\approx \Gamma_0^{-1}$.

The correlation function $C(t)$ in Eq.~\eqref{eq:C(t)} depends on two time scales. The ultraviolet structure of $\nu(\epsilon)$ produces a
short time scale of order $D^{-1}$ and the Fermi factor
introduces a parametrically longer thermal time scale of order
$(\pi T)^{-1}$.
The latter controls the long-time tail
of the correlation function and can dominate the
moment-defined correlation time $\tau_B$ in
Eq.~\eqref{eq:tauB}. Hence, $\tau_B = {\cal O}\left(1/\pi T\right)$,
although the numerical prefactor is not universal because
Eq.~\eqref{eq:tauB} involves $|C(t)|$ and retains some
sensitivity to the ultraviolet line shape.
Since $C(t)$ is proportional to $\Gamma_0$, its overall
normalization cancels from Eq.~\eqref{eq:tauB} but not from~\eqref{eq:tauD}. Hence, $\tau_D^{-1}\propto\Gamma_0\propto|w|^2$, so that $\tau_D$ becomes
parametrically long in the weak-coupling limit.

In Fig.~\ref{fig:bath_timescales}, we plot the numerically calculated $\tau_{B}$, $\tau_{D}$ and $\gamma^{-1}(U)$ as functions of temperature for  $U=100$~mK, and $\Gamma_{0}=0.2$~mK.
The two sets of curves, in solid and dashed lines, are for $D=1.0$~K, and $D=0.5$~K respectively.
In both cases, we see a clear separation of timescales $\tau_{D} \gg \tau_{B}$, which increases with temperature, as expected.
Moreover, $\gamma^{-1}(U) \approx \Gamma_0 \approx  \tau^{-1}_{D}$.
The variation of $\gamma(\omega)$ around $\omega=U$ is within a few percent in an interval $U \pm 50$ mK.

\section{Derivation of the master equation in the secular regime}
\label{app:secular}

The derivation of the secular master equations uses the following consequences of the simple model adopted in the beginning of Sec.~\ref{sec:secular}.
First,  $\epsilon_1=\epsilon_2=\epsilon$.
Second, the odd-parity sector is doubly degenerate, while the energy splitting in the even-parity sector is $2\epsilon$.
Third, the odd-to-even transition frequencies are $\omega_1=\omega_2=U-\epsilon$ and $\omega_3=\omega_4=U+\epsilon$.
Fourth, the transition frequency differences $\delta_{ss'}$ split in two groups: The latter group includes the case with $s=s'$ as well as $(s,s')
\in \{(1,2),(3,4),(2,1),(4,3)\}$. Fifth, the computational Majoranas $a_i$ coincide with the eigenmodes $\tilde{a}_i$.

After all these considerations, and omitting all Lamb shift corrections proportional to $S(\omega)$, the Bloch-Redfield equation simplifies to:
\begin{widetext}
\begin{align}
\nonumber
\dot{\rho} =
&-\frac{1}{8}\gamma(U-\epsilon)\sum_i(\{L^\dagger_{i,1}L_{i,1},\rho\} - 2L_{i,1}\, \rho\, L^{\dagger}_{i,1})
-\frac{1}{8}\gamma(\epsilon-U)\sum_i (\{L_{i,1}L^\dagger_{i,1},\rho\} - 2L^\dagger_{i,1}\, \rho\, L_{i,1})\\
&-\frac{1}{8}\gamma(U+\epsilon)\sum_i (\{L^\dagger_{i,2}L_{i,2},\rho\} - 2L_{i,2}\, \rho\, L^{\dagger}_{i,2})
 -\frac{1}{8}\gamma(-U-\epsilon)\sum_i (\{L_{i,2}L^\dagger_{i,2},\rho\} - 2L^\dagger_{i,2}\, \rho\, L_{i,2})
\end{align}
\end{widetext}
Here, we have introduced the jump operators
\begin{align}
L_{i,1}&=A_{i,1}+A_{i,2}=\Pi_{--}\,a_i\,,\\
L_{i,2}&=A_{i,3}+A_{i,4}=\Pi_{++}\,a_i\,.
\end{align}
Using Eq.~\eqref{eq:jump_operators} and \eqref{eq:projectors} it is possible to show that
\begin{subequations}
\begin{equation}
\sum_i L^\dagger_{i,1}L_{i,1} = \sum_i L_{i,2}L^\dagger_{i,2} = \frac{1}{2}(1-P)
\end{equation}
and
\begin{align}
\sum_i L_{i,1}L^\dagger_{i,1} &= 2\Pi_{--}\,,\\
\sum_i L_{i,2}L^\dagger_{i,2} &= 2\Pi_{++}\,.
\end{align}
\end{subequations}
Performing the sums over $i$, we obtain the master equation in Eq.~\eqref{eq:secular_general_me} of the main text.

\section{Reduction of the Universal Lindblad Equation}
\label{app:ULE}

In this Appendix we show explicitly that the ULE dissipator
in Eq.~\eqref{eq:ULE_dissipator} reduces to the secular
dissipator derived in Sec.~VI for the symmetric geometry
$A_{12}=A_{34}=\epsilon$. We assume disconnected and identical
reservoirs and restrict the dynamics to density matrices satisfying
\begin{equation}
    [\rho,P]=0 .
    \label{eq:app_parity_diag}
\end{equation}
For compactness, we define
\begin{equation}
    L_i\equiv L_{i,1}=\Pi_{--}a_i,
    \qquad
    M_i\equiv L_{i,2}=\Pi_{++}a_i ,
    \label{eq:app_LM}
\end{equation}
and introduce
\begin{align}
    \gamma_- &= \gamma(U-\epsilon),
    &
    \bar{\gamma}_- &= \gamma(\epsilon-U),
    \nonumber\\
    \gamma_+ &= \gamma(U+\epsilon),
    &
    \bar{\gamma}_+ &= \gamma(-U-\epsilon).
    \label{eq:app_rates}
\end{align}
The ULE jump operator in Eq.~\eqref{eq:ULE_jump_special} is then
\begin{equation}
    K_i =
    \sqrt{\gamma_-}\,L_i
    +\sqrt{\bar{\gamma}_-}\,L_i^\dagger
    +\sqrt{\gamma_+}\,M_i
    +\sqrt{\bar{\gamma}_+}\,M_i^\dagger .
    \label{eq:app_K}
\end{equation}

We first consider the anticommutator contribution to the
dissipator. Since both $\Pi_{--}$ and $\Pi_{++}$ project onto
states of even total parity, while each $a_i$ changes the total
parity,
\begin{equation}
    \Pi_{\sigma\sigma}a_i\Pi_{\sigma'\sigma'}=0 ,
    \qquad \sigma,\sigma'=\pm 1 .
    \label{eq:app_parity_selection}
\end{equation}
Together with $\Pi_{--}\Pi_{++}=0$, this implies
\begin{equation}
    L_i^2=M_i^2=L_iM_i=M_iL_i=0 ,
    \label{eq:app_zero_products1}
\end{equation}
as well as the corresponding relations obtained by Hermitian
conjugation. Moreover,
\begin{equation}
    L_i^\dagger M_i=M_i^\dagger L_i
    =L_iM_i^\dagger=M_iL_i^\dagger=0 .
    \label{eq:app_zero_products2}
\end{equation}
All cross terms in $K_i^\dagger K_i$ therefore vanish identically,
and one obtains
\begin{align}
    K_i^\dagger K_i
    ={}&
    \gamma_- L_i^\dagger L_i
    +\bar{\gamma}_- L_i L_i^\dagger
    \nonumber\\
    &+
    \gamma_+ M_i^\dagger M_i
    +\bar{\gamma}_+ M_i M_i^\dagger .
    \label{eq:app_KdagK}
\end{align}

We next consider the recycling term. Expanding
$K_i\rho K_i^\dagger$ gives the four diagonal contributions
\begin{align}
    \left.K_i\rho K_i^\dagger\right|_{\rm diag}
    ={}&
    \gamma_- L_i\rho L_i^\dagger
    +\bar{\gamma}_- L_i^\dagger\rho L_i
    \nonumber\\
    &+
    \gamma_+ M_i\rho M_i^\dagger
    +\bar{\gamma}_+ M_i^\dagger\rho M_i ,
    \label{eq:app_recycling_diag}
\end{align}
together with interference terms. For a density matrix obeying
Eq.~\eqref{eq:app_parity_diag}, terms containing two operators
with the same parity-changing orientation vanish. In particular,
\begin{align}
    L_i\rho L_i
    &=M_i\rho M_i
    =L_i\rho M_i
    =M_i\rho L_i=0 ,
    \nonumber\\
    L_i^\dagger\rho L_i^\dagger
    &=M_i^\dagger\rho M_i^\dagger
    =L_i^\dagger\rho M_i^\dagger
    =M_i^\dagger\rho L_i^\dagger=0 .
    \label{eq:app_parity_cross}
\end{align}
The only remaining interference terms are therefore
\begin{equation}
    L_i\rho M_i^\dagger,\quad
    M_i\rho L_i^\dagger,\quad
    L_i^\dagger\rho M_i,\quad
    M_i^\dagger\rho L_i .
    \label{eq:app_remaining_cross}
\end{equation}
Unlike the terms in Eq.~\eqref{eq:app_parity_cross}, these do not
vanish by total-parity selection alone. Their cancellation follows
from the relative phases with which the two Majoranas of each
canonical pair act on the eigenstates.

Indeed, the eigenstate projectors can be written as
\begin{equation}
    \Pi_{\sigma\tau}
    =
    \frac{1}{4}
    (1-\sigma P_{12})(1-\tau P_{34}),
    \qquad \sigma,\tau=\pm1 ,
    \label{eq:app_projectors}
\end{equation}
and the Majorana operators within each pair satisfy
\begin{equation}
    a_2\Pi_{\sigma\tau}
    =-i\sigma\,a_1\Pi_{\sigma\tau},
    \qquad
    a_4\Pi_{\sigma\tau}
    =-i\tau\,a_3\Pi_{\sigma\tau}.
    \label{eq:app_pair_identity}
\end{equation}
Equivalently,
\begin{equation}
    \Pi_{\sigma\tau}a_2
    =i\sigma\,\Pi_{\sigma\tau}a_1,
    \qquad
    \Pi_{\sigma\tau}a_4
    =i\tau\,\Pi_{\sigma\tau}a_3 .
    \label{eq:app_pair_identity_left}
\end{equation}

Consider, for example, the interference term
\begin{equation}
    \sum_i L_i\rho M_i^\dagger
    =
    \sum_i
    \Pi_{--}a_i\rho a_i\Pi_{++}.
    \label{eq:app_cross_example}
\end{equation}
For the first Majorana pair, Eqs.~\eqref{eq:app_pair_identity}
and \eqref{eq:app_pair_identity_left} give
\begin{align}
    \Pi_{--}a_2\rho a_2\Pi_{++}
    &=
    -\Pi_{--}a_1\rho a_1\Pi_{++},
\end{align}
and therefore
\begin{equation}
    \sum_{i=1}^{2}
    \Pi_{--}a_i\rho a_i\Pi_{++}=0 .
    \label{eq:app_cancel12}
\end{equation}
The same argument applied to the second Majorana pair gives
\begin{equation}
    \sum_{i=3}^{4}
    \Pi_{--}a_i\rho a_i\Pi_{++}=0 .
    \label{eq:app_cancel34}
\end{equation}
Hence
\begin{equation}
    \sum_i L_i\rho M_i^\dagger=0 .
    \label{eq:app_cancel_LMdag}
\end{equation}
Hermitian conjugation gives
\begin{equation}
    \sum_i M_i\rho L_i^\dagger=0 .
    \label{eq:app_cancel_MLdag}
\end{equation}
The two remaining terms cancel in the same way,
\begin{equation}
    \sum_i L_i^\dagger\rho M_i
    =
    \sum_i M_i^\dagger\rho L_i
    =0 .
    \label{eq:app_cancel_reverse}
\end{equation}

Combining Eqs.~\eqref{eq:app_recycling_diag}--%
\eqref{eq:app_cancel_reverse}, the recycling term reduces to
\begin{align}
    \sum_i K_i\rho K_i^\dagger
    ={}&
    \gamma_-\sum_i L_i\rho L_i^\dagger
    +\bar{\gamma}_-\sum_i L_i^\dagger\rho L_i
    \nonumber\\
    &+
    \gamma_+\sum_i M_i\rho M_i^\dagger
    +\bar{\gamma}_+\sum_i M_i^\dagger\rho M_i .
    \label{eq:app_recycling_final}
\end{align}
Substitution of Eqs.~\eqref{eq:app_KdagK} and
\eqref{eq:app_recycling_final} into
Eq.~\eqref{eq:ULE_dissipator} yields
\begin{align}
\mathcal{D}_{\rm ULE}[\rho]
={}&
-\frac{\gamma_-}{8}
\sum_i
\left(
    \{L_i^\dagger L_i,\rho\}
    -2L_i\rho L_i^\dagger
\right)
\nonumber\\
&-\frac{\bar{\gamma}_-}{8}
\sum_i
\left(
    \{L_iL_i^\dagger,\rho\}
    -2L_i^\dagger\rho L_i
\right)
\nonumber\\
&-\frac{\gamma_+}{8}
\sum_i
\left(
    \{M_i^\dagger M_i,\rho\}
    -2M_i\rho M_i^\dagger
\right)
\nonumber\\
&-\frac{\bar{\gamma}_+}{8}
\sum_i
\left(
    \{M_iM_i^\dagger,\rho\}
    -2M_i^\dagger\rho M_i
\right).
    \label{eq:app_ULE_reduced}
\end{align}
Upon restoring the definitions in
Eq.~\eqref{eq:app_rates}, Eq.~\eqref{eq:app_ULE_reduced} is
identical to the dissipative part of the master equation in the secular regime. 

\bibliography{references}

@article{dvir2023,
  title={Realization of a minimal {Kitaev} chain in coupled quantum dots},
  author={Dvir, Tom and Wang, Guanzhong and van Loo, Nick and Liu, Chun-Xiao and Mazur, Grzegorz P and Bordin, Alberto and Ten Haaf, Sebastiaan LD and Wang, Ji-Yin and van Driel, David and Zatelli, Francesco and Li, Xiang and Malinowski, Filip K and Gazibegovic, Sasa and Badawy, Ghada and Bakkers, Erik P. A. M. and Wimmer, Michael and Kouwenhoven, Leo P},
  journal={Nature},
  volume={614},
  number={7948},
  pages={445--450},
  year={2023},
  doi={https://doi.org/10.1038/s41586-022-05585-1},
  publisher={Nature Publishing Group UK London}
}

@article{pedrocchi2015,
  title = {Majorana Braiding with Thermal Noise},
  author = {Pedrocchi, Fabio L. and DiVincenzo, David P.},
  journal = {Phys. Rev. Lett.},
  volume = {115},
  issue = {12},
  pages = {120402},
  numpages = {6},
  year = {2015},
  month = {Sep},
  publisher = {American Physical Society},
  doi = {10.1103/PhysRevLett.115.120402},
  url = {https://link.aps.org/doi/10.1103/PhysRevLett.115.120402}
}

@article{lai2020,
  title = {Decoherence dynamics of {Majorana} qubits under braiding operations},
  author = {Lai, Hon-Lam and Zhang, Wei-Min},
  journal = {Phys. Rev. B},
  volume = {101},
  issue = {19},
  pages = {195428},
  numpages = {10},
  year = {2020},
  month = {May},
  publisher = {American Physical Society},
  doi = {10.1103/PhysRevB.101.195428},
  url = {https://link.aps.org/doi/10.1103/PhysRevB.101.195428}
}

@misc{sahu2026,
      title={Noisy Braiding of {Majorana} Modes: A Comparison of Nanowire Trijunction and Quantum-Dot-Assisted Architectures}, 
      author={Dibyajyoti Sahu and Suhas Gangadharaiah},
      year={2026},
      eprint={2608.09416},
      archivePrefix={arXiv},
      primaryClass={cond-mat.mes-hall},
      url={https://arxiv.org/abs/2608.09416}, 
}

@article{ten2025,
  title={Observation of edge and bulk states in a three-site {Kitaev} chain},
  author={Ten Haaf, Sebastiaan LD and Zhang, Yining and Wang, Qingzhen and Bordin, Alberto and Liu, Chun-Xiao and Kulesh, Ivan and Sietses, Vincent PM and Prosko, Christian G and Xiao, Di and Thomas, Candice and Manfra, Michael J and Wimmer, Michael and Goswami, Srijit},
  journal={Nature},
  volume={641},
  number={8064},
  pages={890--895},
  year={2025},
  doi={https://doi.org/10.1038/s41586-025-08892-5},
  publisher={Nature Publishing Group UK London}
}

@article{vanloo2026,
  title={Single-shot parity readout of a minimal {Kitaev} chain},
  author={van Loo, Nick and Zatelli, Francesco and Steffensen, Gorm O and Roovers, Bart and Wang, Guanzhong and Van Caekenberghe, Thomas and Bordin, Alberto and van Driel, David and Zhang, Yining and Huisman, Wietze D and Badawy, Ghada and Bakkers, Erik P. A. M. and Mazue, Grzegorz P. and Aguado, Ramón and Kouwenhoven, Leo P.},
  journal={Nature},
  volume={650},
  number={8101},
  pages={334--339},
  year={2026},
  doi={https://doi.org/10.1038/s41586-025-09927-7},
  publisher={Nature Publishing Group UK London}
}

@article{sau2012,
  title={Realizing a robust practical {Majorana chain} in a quantum-dot-superconductor linear array},
  author={Sau, Jay D and Das Sarma, S},
  journal={Nature Communications},
  volume={3},
  number={1},
  pages={964},
  year={2012},
  doi={https://doi.org/10.1038/ncomms1966},
  publisher={Nature Publishing Group UK London}
}

@article{microsoft2025,
  title={Interferometric single-shot parity measurement in {InAs--Al} hybrid devices},
  author={Aghaee, Morteza and Akkala, Arun and Alam, Zulfi and Ali, Rizwan and Alcaraz Ramirez, Alejandro and Andrzejczuk, Mariusz and Antipov, Andrey E. and Aseev, Pavel and Astafev, Mikhail and Bauer, Bela and Becker, Jonathan and Boddapati, Srini and Boekhout, Frenk and Bommer, Jouri and Bosma, Tom and Bourdet, Leo and Boutin, Samuel and Caroff, Philippe and Casparis, Lucas and Cassidy, Maja and Chatoor, Sohail and Christensen, Anna Wulf and Clay, Noah and Cole, William S. and Corsetti, Fabiano and Cui, Ajuan and Dalampiras, Paschalis and Dokania, Anand and de Lange, Gijs and de Moor, Michiel and Estrada Saldaña, Juan Carlos and Fallahi, Saeed and Fathabad, Zahra Heidarnia and Gamble, John and Gardner, Geoff and Govender, Deshan and Griggio, Flavio and Grigoryan, Ruben and Gronin, Sergei and Gukelberger, Jan and Hansen, Esben Bork and Heedt, Sebastian and Herranz Zamorano, Jesús and Ho, Samantha and Holgaard, Ulrik Laurens and Ingerslev, Henrik and Johansson, Linda and Jones, Jeffrey and Kallaher, Ray and Karimi, Farhad and Karzig, Torsten and King, Evelyn and Kloster, Maren Elisabeth and Knapp, Christina and Kocon, Dariusz and Koski, Jonne and Kostamo, Pasi and Krogstrup, Peter and Kumar, Mahesh and Laeven, Tom and Larsen, Thorvald and Li, Kongyi and Lindemann, Tyler and Love, Julie and Lutchyn, Roman and Madsen, Morten Hannibal and Manfra, Michael and Markussen, Signe and Martinez, Esteban and McNeil, Robert and Memisevic, Elvedin and Morgan, Trevor and Mullally, Andrew and Nayak, Chetan and Nielsen, Jens and Nielsen, William Hvidtfelt Padkær and Nijholt, Bas and Nurmohamed, Anne and O’Farrell, Eoin and Otani, Keita and Pauka, Sebastian and Petersson, Karl and Petit, Luca and Pikulin, Dmitry I. and Preiss, Frank and Quintero-Perez, Marina and Rajpalke, Mohana and Rasmussen, Katrine and Razmadze, Davydas and Reentila, Outi and Reilly, David and Rouse, Richard and Sadovskyy, Ivan and Sainiemi, Lauri and Schreppler, Sydney and Sidorkin, Vadim and Singh, Amrita and Singh, Shilpi and Sinha, Sarat and Sohr, Patrick and Stankevič, Tomaš and Stek, Lieuwe and Suominen, Henri and Suter, Judith and Svidenko, Vicky and Teicher, Sam and Temuerhan, Mine and Thiyagarajah, Nivetha and Tholapi, Raj and Thomas, Mason and Toomey, Emily and Upadhyay, Shivendra and Urban, Ivan and Vaitiekėnas, Saulius and Van Hoogdalem, Kevin and Van Woerkom, David and Viazmitinov, Dmitrii V. and Vogel, Dominik and Waddy, Steven and Watson, John and Weston, Joseph and Winkler, Georg W. and Yang, Chung Kai and Yau, Sean and Yi, Daniel and Yucelen, Emrah and Webster, Alex and Zeisel, Roland and Zhao, Ruichen},
  journal={Nature},
  volume={638},
  number={8051},
  pages={651--655},
  year={2025},
  doi={https://doi.org/10.1038/s41586-024-08445-2},
  publisher={Nature Publishing Group UK London}
}

@article{dassarma2023,
  title={In search of {Majorana}},
  author={Das Sarma, Sankar},
  journal={Nature Physics},
  volume={19},
  number={2},
  pages={165--170},
  year={2023},
  doi={https://doi.org/10.1038/s41567-022-01900-9},
  publisher={Nature Publishing Group UK London}
}

@article{legg2026,
  title={On the robustness of topological gap detection via transport},
  author={Legg, Henry F},
  journal={Nature},
  volume={654},
  number={8120},
  pages={E22--E26},
  year={2026},
  doi={https://doi.org/10.1038/s41586-026-10567-8},
  publisher={Nature Publishing Group UK London}
}

@article{microsoft2026reply,
  author={{Microsoft Quantum}},
  title={Reply to: {On} the robustness of topological gap detection via transport},
  journal={Nature},
  volume={654},
  number={8120},
  pages={E27--E28},
  year={2026},
  doi={https://doi.org/10.1038/s41586-026-10568-7},
  publisher={Nature Publishing Group UK London}
}

@misc{aghaee2025,
      title={Distinct Lifetimes for {$X$} and {$Z$} Loop Measurements in a {Majorana} Tetron Device}, 
      author={Morteza Aghaee and Zulfi Alam and Rikke Andersen and Mariusz Andrzejczuk and Andrey Antipov and Mikhail Astafev and Lukas Avilovas and Ahmad Azizimanesh and Eric Banek and Bela Bauer and Jonathan Becker and Umesh Kumar Bhaskar and Andrea G. Boa and Srini Boddapati and Nichlaus Bohac and Jouri D. S. Bommer and Jan Borovsky and Léo Bourdet and Samuel Boutin and Lucas Casparis and Srivatsa Chakravarthi and Hamidreza Chalabi and Benjamin J. Chapman and Nikolaos Chatzaras and Tzu-Chiao Chien and Jason Cho and Patrick Codd and William Cole and Paul W. Cooper and Fabiano Corsetti and Ajuan Cui and Tareq El Dandachi and Celine Dinesen and Andreas Ekefjärd and Saeed Fallahi and Luca Galletti and Geoffrey C. Gardner and Gonzalo Leon Gonzalez and Deshan Govender and Flavio Griggio and Ruben Grigoryan and Sebastian Grijalva and Sergei Gronin and Jan Gukelberger and Marzie Hamdast and A. Ben Hamida and Esben Bork Hansen and Caroline Tynell Hansen and Sebastian Heedt and Samantha Ho and Laurens Holgaard and Kevin van Hoogdalem and John Hornibrook and Henrik Ingerslev and Lovro Ivancevic and Sherwan Jamo and Max Jantos and Thomas Jensen and Jaspreet Singh Jhoja and Jeffrey C Jones and Vidul Joshi and Konstantin V. Kalashnikov and Ray Kallaher and Rachpon Kalra and Farhad Karimi and Torsten Karzig and Seth Kimes and Evelyn King and Maren Elisabeth Kloster and Christina Knapp and Jonne V. Koski and Pasi Kostamo and Tom Laeven and Jeffrey Lai and Gijs de Lange and Thorvald W. Larsen and Kyunghoon Lee and Kongyi Li and Guangze Li and Shuang Liang and Tyler Lindemann and Matthew Looij and Marijn Lucas and Roman Lutchyn and Morten Hannibal Madsen and Nasiari Madulid and Michael J. Manfra and Laveena Manjunath and Signe Markussen and Esteban Martinez and Marco Mattila and J. R. Mattinson and R. P. G. McNeil and Alba Pérez Millan and Ryan V. Mishmash and Sarang Mittal and Christian Møllgaard and M. W. A. de Moor and Eduardo Puchol Morejon and Trevor Morgan and George Moussa and B. P. Nabar and Anirudh Narla and Chetan Nayak and Jens Hedegaard Nielsen and William Hvidtfelt Padkær Nielsen and Frédéric Nolet and Michael J. Nystrom and Eoin O'Farrell and Thomas A. Ohki and Keita Otani and Camille Papon and Karl D Petersson and Luca Petit and Dima Pikulin and Mohana Rajpalke and Alejandro Alcaraz Ramirez and David Razmadze and Yuan Ren and Ivan Sadovskyy and Lauri Sainiemi and Juan Carlos Estrada Saldaña and Irene Sanlorenzo and Tatiane Pereira dos Santos and Simon Schaal and John Schack and Emma R. Schmidgall and Christina Sfetsou and Cristina Sfiligoj and Sarat Sinha and Patrick Sohr and Thomas L. Sørensen and Kasper Spiegelhauer and Tomaš Stankević and Lieuwe J. Stek and Patrick Strøm-Hansen and Henri J. Suominen and Judith Suter and Samuel M. L. Teicher and Raj Tholapi and Mason Thomas and D. W. Tom and Emily Toomey and Joshua Tracy and Michelle Turley and Matthew D. Turner and Shivendra Upadhyay and Ivan Urban and Dmitrii V. Viazmitinov and Anna Wulff Viazmitinova and Beatriz Viegas and Dominik J. Vogel and John Watson and Alex Webster and Joseph Weston and Timothy Williamson and Georg W. Winkler and David J. van Woerkom and Brian Paquelet Wuetz and Chung-Kai Yang and Shang-Jyun and Yu and Emrah Yucelen and Jesús Herranz Zamorano and Roland Zeisel and Guoji Zheng and A. M. Zimmerman},
      year={2025},
      eprint={2507.08795},
      archivePrefix={arXiv},
      primaryClass={cond-mat.mes-hall},
      url={https://arxiv.org/abs/2507.08795}, 
}

@article{munk2020,
  title = {Parity-to-charge conversion in {Majorana} qubit readout},
  author = {Munk, Morten I. K. and Schulenborg, Jens and Egger, Reinhold and Flensberg, Karsten},
  journal = {Phys. Rev. Res.},
  volume = {2},
  issue = {3},
  pages = {033254},
  numpages = {19},
  year = {2020},
  month = {Aug},
  publisher = {American Physical Society},
  doi = {10.1103/PhysRevResearch.2.033254},
  url = {https://link.aps.org/doi/10.1103/PhysRevResearch.2.033254}
}

@misc{boutin2025,
      title={Predictive simulations of the dynamical response of mesoscopic devices}, 
      author={Samuel Boutin and Torsten Karzig and Tareq El Dandachi and Ryan V. Mishmash and Jan Gukelberger and Roman M. Lutchyn and Bela Bauer},
      year={2025},
      eprint={2502.12960},
      archivePrefix={arXiv},
      primaryClass={cond-mat.mes-hall},
      url={https://arxiv.org/abs/2502.12960}, 
}

@misc{aghaee2026,
      title={20 Second Parity Lifetime in an {InAs--Pb} Tetron Device}, 
      author={Morteza Aghaee and Zulfi Alam and Mariusz Andrzejczuk and Andrey Antipov and Theodora Asimakidis and Mikhail Astafev and Lukas Avilovas and Ahmad Azizimanesh and Amin Barzegar and Bela Bauer and Jonathan Becker and Umesh Kumar Bhaskar and Andrea G. Boa and Srini Boddapati and Nichlaus Bohac and Jouri Bommer and Jan Borovsky and Léo Bourdet and Samuel Boutin and Srivatsa Chakravarthi and Benjamin J. Chapman and Nikolaos Chatzaras and Tzu-Chiao Chien and Jason Cho and Patrick T. Codd and William Cole and Paul W. Cooper and Fabiano Corsetti and Ajuan Cui and Tareq El Dandachi and Konstantinos Divanis and Clayton Doyle and Andreas Ekefjard and Javier A. Falcon and Saeed Fallahi and Luca Galletti and Geoffrey C. Gardner and Haris Gavranovic and João Pedro Morais Gomes and Deshan Govender and Flavio Griggio and Ruben Grigoryan and Sebastian Grijalva and Sergei Gronin and Jan Gukelberger and Marzie Hamdast and Esben Bork Hansen and Sebastian Heedt and Samantha Ho and Laurens Holgaard and Kevin van Hoogdalem and Jinnapat Indrapiromkul and Henrik Ingerslev and Lovro Ivancevic and Max Jantos and Thomas Jensen and Jaspreet Singh Jhoja and Vidul R. Joshi and Konstantin V. Kalashnikov and Ray Kallaher and Rachpon Kalra and Farhad Karimi and Torsten Karzig and Maren Elisabeth Kloster and Christina Knapp and Jonathan Knoblauch and Jonne Koski and Anders Kringhøj and Tom Laeven and Jeffrey Lai and Gijs de Lange and Thorvald W. Larsen and Kyunghoon Lee and Kongyi Li and Shuang Liang and Tyler Lindemann and Luna Lochmatter and Marijn Lucas and Roman Lutchyn and Morten Hannibal Madsen and Nasiari Madulid and Ivan Maliyov and Yanick Mampaey and Michael Manfra and Signe Brynold Markussen and Esteban A. Martinez and J. R. Mattinson and Mónica Meira and Camille A. Mikolas and Sarang Mittal and Gopakumar Mohandas and Christian Mollgaard and Michiel W. A. de Moor and Chris Moore and George Moussa and Bhargav Nabar and Anirudh Narla and Ahmad Naseri and Chetan Nayak and Bjørn Funch Schrøder Nielsen and Jens Hedegaard Nielsen and Michael J. Nystrom and Eoin O'Farrell and Keita Ohtani and Theodore Rex Orth and Sara Di Paolo and Camille Papon and Luca Petit and Dima Pikulin and Mohana Rajpalke and Alejandro Alcaraz Ramirez and Katrine Rasmussen and David Razmadze and Yuan Ren and Mariya Romanova and Loïc Roure and Ivan Sadovskyy and Lauri Sainiemi and Juan Carlos Estrada Saldaña and Irene Sanlorenzo and Tatiane Pereira dos Santos and Carl Vincent C. Saruda and Simon Schaal and John Schack and Emma R. Schmidgall and Christina Sfetsou and Cristina Sfiligoj and Zahra Shekason and Sarat Shankar Sinha and Patrick Sohr and Maria João Lourenço de Sousa and Kasper Roed Spiegelhauer and Tomas Stankevic and Henri J. Suominen and Judith Suter and Attila Szénási and Samuel M. L. Teicher and Naganivetha Thiyagarajah and Raj Tholapi and Mason Thomas and Dennis Tom and Emily Toomey and Joshua Tracy and Michelle Turley and Matthew D. Turner and Ivan Urban and Aakash Valliappan and Dmitrii V. Viazmitinov and Anna Wulff Viazmitinova and Dominik Johannes Vogel and Wenbo Wang and Christopher A. Watson and John Watson and Alex Webster and Joseph Weston and Timothy Williamson and Georg W. Winkler and David J. van Woerkom and Brian Paquelet Wuetz and Cécile X. Yu and Emrah Yucelen and Jesús Herranz Zamorano and Roland Zeisel and Guoji Zheng and A. M. Zimmerman},
      year={2026},
      eprint={2606.03884},
      archivePrefix={arXiv},
      primaryClass={cond-mat.mes-hall},
      url={https://arxiv.org/abs/2606.03884}, 
}

@article{nitsch2025,
  title = {Poor Man's {Majorana} Tetron},
  author = {Nitsch, Maximilian and Maffi, Lorenzo and Baran, Virgil V. and Seoane Souto, Rub\'en and Paaske, Jens and Leijnse, Martin and Burrello, Michele},
  journal = {PRX Quantum},
  volume = {6},
  issue = {3},
  pages = {030365},
  numpages = {32},
  year = {2025},
  month = {Sep},
  publisher = {American Physical Society},
  doi = {10.1103/r75t-jv32},
  url = {https://link.aps.org/doi/10.1103/r75t-jv32}
}

@article{pan2025,
  title = {{Rabi and Ramsey} oscillations of a {Majorana} qubit in a quantum dot-superconductor array},
  author = {Pan, Haining and Das Sarma, Sankar and Liu, Chun-Xiao},
  journal = {Phys. Rev. B},
  volume = {111},
  issue = {7},
  pages = {075416},
  numpages = {12},
  year = {2025},
  month = {Feb},
  publisher = {American Physical Society},
  doi = {10.1103/PhysRevB.111.075416},
  url = {https://link.aps.org/doi/10.1103/PhysRevB.111.075416}
}

@misc{zatelli2026,
      title={Majorana parity qubit in coupled minimal {Kitaev} chains}, 
      author={Francesco Zatelli and Bart Roovers and Nick van Loo and Antonio Lombardi and Juan D. Torres Luna and Sebastian Miles and Vincent P. M. Sietses and Florian J. Bennebroek Evertsz' and Pablo Cova Fariña and Alberto Bordin and Ghada Badawy and Erik P. A. M. Bakkers and Michael Wimmer and Leo P. Kouwenhoven},
      year={2026},
      eprint={2607.09511},
      archivePrefix={arXiv},
      primaryClass={cond-mat.mes-hall},
      url={https://arxiv.org/abs/2607.09511}, 
}

@article{aasen2016,
  title = {Milestones Toward {Majorana-Based} Quantum Computing},
  author = {Aasen, David and Hell, Michael and Mishmash, Ryan V. and Higginbotham, Andrew and Danon, Jeroen and Leijnse, Martin and Jespersen, Thomas S. and Folk, Joshua A. and Marcus, Charles M. and Flensberg, Karsten and Alicea, Jason},
  journal = {Phys. Rev. X},
  volume = {6},
  issue = {3},
  pages = {031016},
  numpages = {28},
  year = {2016},
  month = {Aug},
  publisher = {American Physical Society},
  doi = {10.1103/PhysRevX.6.031016},
  url = {https://link.aps.org/doi/10.1103/PhysRevX.6.031016}
}

@article{mozgunov2020,
  title={Completely positive master equation for arbitrary driving and small level spacing},
  author={Mozgunov, Evgeny and Lidar, Daniel},
  journal={Quantum},
  volume={4},
  pages={227},
  year={2020},
  doi={https://doi.org/10.22331/q-2020-02-06-227},
  publisher={Verein zur F{\"o}rderung des Open Access Publizierens in den Quantenwissenschaften}
}

@article{vijay2016,
  title = {Teleportation-based quantum information processing with {Majorana} zero modes},
  author = {Vijay, Sagar and Fu, Liang},
  journal = {Phys. Rev. B},
  volume = {94},
  issue = {23},
  pages = {235446},
  numpages = {9},
  year = {2016},
  month = {Dec},
  publisher = {American Physical Society},
  doi = {10.1103/PhysRevB.94.235446},
  url = {https://link.aps.org/doi/10.1103/PhysRevB.94.235446}
}

@book{breuer2002,
  title={The theory of open quantum systems},
  author={Breuer, Heinz-Peter and Petruccione, Francesco},
  year={2002},
  publisher={OUP Oxford}
}

@article{bhattacharyya2026,
  title = {Decoherence of {Majorana} zero modes mediated by gapless fermions},
  author = {Bhattacharyya, Sauri and Grilli, Marco and van Heck, Bernard},
  journal = {Phys. Rev. B},
  volume = {113},
  issue = {3},
  pages = {035422},
  numpages = {19},
  year = {2026},
  month = {Jan},
  publisher = {American Physical Society},
  doi = {10.1103/hfkz-jptt},
  url = {https://link.aps.org/doi/10.1103/hfkz-jptt}
}

@article{flensberg2021,
  title={Engineered platforms for topological superconductivity and {Majorana} zero modes},
  author={Flensberg, Karsten and von Oppen, Felix and Stern, Ady},
  journal={Nature Reviews Materials},
  volume={6},
  number={10},
  pages={944--958},
  year={2021},
  doi={https://doi.org/10.1038/s41578-021-00336-6},
  publisher={Nature Publishing Group UK London}
}

@misc{alase2025,
      title={Decoherence of {Majorana} qubits by 1/f noise}, 
      author={Abhijeet Alase and Marcus C. Goffage and Maja C. Cassidy and Susan N. Coppersmith},
      year={2025},
      eprint={2506.22394},
      archivePrefix={arXiv},
      primaryClass={cond-mat.mes-hall},
      url={https://arxiv.org/abs/2506.22394}, 
}

@article{nayak2008,
  title = {{Non-Abelian} anyons and topological quantum computation},
  author = {Nayak, Chetan and Simon, Steven H. and Stern, Ady and Freedman, Michael and Das Sarma, Sankar},
  journal = {Rev. Mod. Phys.},
  volume = {80},
  issue = {3},
  pages = {1083--1159},
  numpages = {0},
  year = {2008},
  month = {Sep},
  publisher = {American Physical Society},
  doi = {10.1103/RevModPhys.80.1083},
  url = {https://link.aps.org/doi/10.1103/RevModPhys.80.1083}
}

@article{dassarma2015,
  title={Majorana zero modes and topological quantum computation},
  author={Das Sarma, Sankar and Freedman, Michael and Nayak, Chetan},
  journal={npj Quantum Information},
  volume={1},
  number={1},
  pages={1--13},
  year={2015},
  doi={https://doi.org/10.1038/npjqi.2015.1},
  publisher={Nature Publishing Group}
}

@article{budich2012,
  title = {Failure of protection of {Majorana} based qubits against decoherence},
  author = {Budich, Jan Carl and Walter, Stefan and Trauzettel, Bj\"orn},
  journal = {Phys. Rev. B},
  volume = {85},
  issue = {12},
  pages = {121405},
  numpages = {4},
  year = {2012},
  month = {Mar},
  publisher = {American Physical Society},
  doi = {10.1103/PhysRevB.85.121405},
  url = {https://link.aps.org/doi/10.1103/PhysRevB.85.121405}
}

@article{rainis2012,
  title = {Majorana qubit decoherence by quasiparticle poisoning},
  author = {Rainis, Diego and Loss, Daniel},
  journal = {Phys. Rev. B},
  volume = {85},
  issue = {17},
  pages = {174533},
  numpages = {10},
  year = {2012},
  month = {May},
  publisher = {American Physical Society},
  doi = {10.1103/PhysRevB.85.174533},
  url = {https://link.aps.org/doi/10.1103/PhysRevB.85.174533}
}

@article{karzig2021,
  title = {Quasiparticle Poisoning of {Majorana} Qubits},
  author = {Karzig, Torsten and Cole, William S. and Pikulin, Dmitry I.},
  journal = {Phys. Rev. Lett.},
  volume = {126},
  issue = {5},
  pages = {057702},
  numpages = {7},
  year = {2021},
  month = {Feb},
  publisher = {American Physical Society},
  doi = {10.1103/PhysRevLett.126.057702},
  url = {https://link.aps.org/doi/10.1103/PhysRevLett.126.057702}
}

@article{goldstein2011,
  title = {Decay rates for topological memories encoded with {Majorana} fermions},
  author = {Goldstein, G. and Chamon, C.},
  journal = {Phys. Rev. B},
  volume = {84},
  issue = {20},
  pages = {205109},
  numpages = {23},
  year = {2011},
  month = {Nov},
  publisher = {American Physical Society},
  doi = {10.1103/PhysRevB.84.205109},
  url = {https://link.aps.org/doi/10.1103/PhysRevB.84.205109}
}

@article{cheng2012,
  title = {Topological protection of {Majorana} qubits},
  author = {Cheng, Meng and Lutchyn, Roman M. and Das Sarma, S.},
  journal = {Phys. Rev. B},
  volume = {85},
  issue = {16},
  pages = {165124},
  numpages = {8},
  year = {2012},
  month = {Apr},
  publisher = {American Physical Society},
  doi = {10.1103/PhysRevB.85.165124},
  url = {https://link.aps.org/doi/10.1103/PhysRevB.85.165124}
}

@article{bonderson2013,
  title = {Quasi-topological phases of matter and topological protection},
  author = {Bonderson, Parsa and Nayak, Chetan},
  journal = {Phys. Rev. B},
  volume = {87},
  issue = {19},
  pages = {195451},
  numpages = {25},
  year = {2013},
  month = {May},
  publisher = {American Physical Society},
  doi = {10.1103/PhysRevB.87.195451},
  url = {https://link.aps.org/doi/10.1103/PhysRevB.87.195451}
}

@article{aasen2025,
  title = {Blueprint for fault-tolerant quantum computation with topological qubit arrays},
  author = {Aasen, David and Aghaee, Morteza and Alam, Zulfi and Andrzejczuk, Mariusz and Antipov, Andrey and Astafev, Mikhail and Avilovas, Lukas and Barzegar, Amin and Bauer, Bela and Becker, Jonathan and Bello-Rivas, Juan M. and Bhaskar, Umesh and Bocharov, Alex and Boddapati, Srini and Bohn, David and Bommer, Jouri and Bonderson, Parsa and Borovsky, Jan and Bourdet, Leo and Boutin, Samuel and Brown, Tom and Campbell, Gary and Casparis, Lucas and Chakravarthi, Srivatsa and Chao, Rui and Chapman, Benjamin J. and Chatoor, Sohail and Christensen, Anna Wulff and Codd, Patrick and Cole, William and Cooper, Paul and Corsetti, Fabiano and Cui, Ajuan and van Dam, Wim and Dandachi, Tareq El and Daraeizadeh, Sahar and Dumitrascu, Adrian and Ekefj\"ard, Andreas and Fallahi, Saeed and Galletti, Luca and Gardner, Geoff and Gatta, Raghu and Gavranovic, Haris and Goulding, Michael and Govender, Deshan and Griggio, Flavio and Grigoryan, Ruben and Grijalva, Sebastian and Gronin, Sergei and Gukelberger, Jan and Haah, Jeongwan and Hamdast, Marzie and Hansen, Esben Bork and Hastings, Matthew and Heedt, Sebastian and Ho, Samantha and Hogaboam, Justin and Holgaard, Laurens and Van Hoogdalem, Kevin and Indrapiromkul, Jinnapat and Ingerslev, Henrik and Ivancevic, Lovro and Jablonski, Sarah and Jensen, Thomas and Jhoja, Jaspreet and Jones, Jeffrey and Kalashnikov, Kostya and Kallaher, Ray and Kalra, Rachpon and Karimi, Farhad and Karzig, Torsten and Kimes, Seth and Kliuchnikov, Vadym and Kloster, Maren Elisabeth and Knapp, Christina and Knee, Derek and Koski, Jonne and Kostamo, Pasi and Kuesel, Jamie and Lackey, Brad and Laeven, Tom and Lai, Jeffrey and de Lange, Gijs and Larsen, Thorvald and Lee, Jason and Lee, Kyunghoon and Leum, Grant and Li, Kongyi and Lindemann, Tyler and Lucas, Marijn and Lutchyn, Roman and Madsen, Morten Hannibal and Madulid, Nash and Manfra, Michael and Markussen, Signe Brynold and Martinez, Esteban and Mattila, Marco and Mattinson, Jake and McNeil, Robert and Mei, Antonio Rodolph and Mishmash, Ryan V. and Mohandas, Gopakumar and Mollgaard, Christian and de Moor, Michiel and Morgan, Trevor and Moussa, George and Narla, Anirudh and Nayak, Chetan and Nielsen, Jens Hedegaard and Nielsen, William Hvidtfelt Padk\ae{}r and Nolet, Fr\'ed\'eric and Nystrom, Mike and O'Farrell, Eoin and Otani, Keita and Paetznick, Adam and Papon, Camille and Paz, Andres and Petersson, Karl and Petit, Luca and Pikulin, Dima and Pons, Diego Olivier Fernandez and Quinn, Sam and Rajpalke, Mohana and Ramirez, Alejandro Alcaraz and Rasmussen, Katrine and Razmadze, David and Reichardt, Ben and Ren, Yuan and Reneris, Ken and Riccomini, Roy and Sadovskyy, Ivan and Sainiemi, Lauri and Salda\~na, Juan Carlos Estrada and Sanlorenzo, Irene and Schaal, Simon and Schmidgall, Emma and Sfiligoj, Cristina and da Silva, Marcus P. and Singh, Shilpi and Sinha, Sarat and Soeken, Mathias and Sohr, Patrick and Stankevic, Tomas and Stek, Lieuwe and Str\o{}m-Hansen, Patrick and Stuppard, Eric and Sundaram, Aarthi and Suominen, Henri and Suter, Judith and Suzuki, Satoshi and Svore, Krysta and Teicher, Sam and Thiyagarajah, Nivetha and Tholapi, Raj and Thomas, Mason and Tom, Dennis and Toomey, Emily and Tracy, Josh and Troyer, Matthias and Turley, Michelle and Turner, Matthew D. and Upadhyay, Shivendra and Urban, Ivan and Vaschillo, Alexander and Viazmitinov, Dmitrii and Vogel, Dominik and Wang, Zhenghan and Watson, John and Webster, Alex and Weston, Joseph and Williamson, Timothy and Winkler, Georg W. and van Woerkom, David J. and W\"utz, Brian Paquelet and Yang, Chung Kai and Yu, Richard and Yucelen, Emrah and Zamorano, Jes\'us Herranz and Zeisel, Roland and Zheng, Guoji and Zilke, Justin and Zimmerman, Andrew},
  journal = {Phys. Rev. Res.},
  volume = {7},
  issue = {4},
  pages = {041002},
  numpages = {24},
  year = {2025},
  month = {Dec},
  publisher = {American Physical Society},
  doi = {10.1103/qx36-4rv1},
  url = {https://link.aps.org/doi/10.1103/qx36-4rv1}
}

@article{aghaee2023,
  title = {{InAs-Al} hybrid devices passing the topological gap protocol},
  author = {Aghaee, Morteza and Akkala, Arun and Alam, Zulfi and Ali, Rizwan and Alcaraz Ramirez, Alejandro and Andrzejczuk, Mariusz and Antipov, Andrey E. and Aseev, Pavel and Astafev, Mikhail and Bauer, Bela and Becker, Jonathan and Boddapati, Srini and Boekhout, Frenk and Bommer, Jouri and Bosma, Tom and Bourdet, Leo and Boutin, Samuel and Caroff, Philippe and Casparis, Lucas and Cassidy, Maja and Chatoor, Sohail and Christensen, Anna Wulf and Clay, Noah and Cole, William S. and Corsetti, Fabiano and Cui, Ajuan and Dalampiras, Paschalis and Dokania, Anand and de Lange, Gijs and de Moor, Michiel and Estrada Salda\~na, Juan Carlos and Fallahi, Saeed and Fathabad, Zahra Heidarnia and Gamble, John and Gardner, Geoff and Govender, Deshan and Griggio, Flavio and Grigoryan, Ruben and Gronin, Sergei and Gukelberger, Jan and Hansen, Esben Bork and Heedt, Sebastian and Herranz Zamorano, Jes\'us and Ho, Samantha and Holgaard, Ulrik Laurens and Ingerslev, Henrik and Johansson, Linda and Jones, Jeffrey and Kallaher, Ray and Karimi, Farhad and Karzig, Torsten and King, Evelyn and Kloster, Maren Elisabeth and Knapp, Christina and Kocon, Dariusz and Koski, Jonne and Kostamo, Pasi and Krogstrup, Peter and Kumar, Mahesh and Laeven, Tom and Larsen, Thorvald and Li, Kongyi and Lindemann, Tyler and Love, Julie and Lutchyn, Roman and Madsen, Morten Hannibal and Manfra, Michael and Markussen, Signe and Martinez, Esteban and McNeil, Robert and Memisevic, Elvedin and Morgan, Trevor and Mullally, Andrew and Nayak, Chetan and Nielsen, Jens and Nielsen, William Hvidtfelt Padk\ae{}r and Nijholt, Bas and Nurmohamed, Anne and O'Farrell, Eoin and Otani, Keita and Pauka, Sebastian and Petersson, Karl and Petit, Luca and Pikulin, Dmitry I. and Preiss, Frank and Quintero-Perez, Marina and Rajpalke, Mohana and Rasmussen, Katrine and Razmadze, Davydas and Reentila, Outi and Reilly, David and Rouse, Richard and Sadovskyy, Ivan and Sainiemi, Lauri and Schreppler, Sydney and Sidorkin, Vadim and Singh, Amrita and Singh, Shilpi and Sinha, Sarat and Sohr, Patrick and Stankevi\ifmmode \check{c}\else \v{c}\fi{}, Toma\ifmmode \check{s}\else \v{s}\fi{} and Stek, Lieuwe and Suominen, Henri and Suter, Judith and Svidenko, Vicky and Teicher, Sam and Temuerhan, Mine and Thiyagarajah, Nivetha and Tholapi, Raj and Thomas, Mason and Toomey, Emily and Upadhyay, Shivendra and Urban, Ivan and Vaitiek\ifmmode \dot{e}\else \.{e}\fi{}nas, Saulius and Van Hoogdalem, Kevin and Van Woerkom, David and Viazmitinov, Dmitrii V. and Vogel, Dominik and Waddy, Steven and Watson, John and Weston, Joseph and Winkler, Georg W. and Yang, Chung Kai and Yau, Sean and Yi, Daniel and Yucelen, Emrah and Webster, Alex and Zeisel, Roland and Zhao, Ruichen},
  collaboration = {Microsoft Quantum},
  journal = {Phys. Rev. B},
  volume = {107},
  issue = {24},
  pages = {245423},
  numpages = {54},
  year = {2023},
  month = {Jun},
  publisher = {American Physical Society},
  doi = {10.1103/PhysRevB.107.245423},
  url = {https://link.aps.org/doi/10.1103/PhysRevB.107.245423}
}

@article{mishmash2020,
  title = {Dephasing and leakage dynamics of noisy {Majorana}-based qubits: {Topological versus Andreev}},
  author = {Mishmash, Ryan V. and Bauer, Bela and von Oppen, Felix and Alicea, Jason},
  journal = {Phys. Rev. B},
  volume = {101},
  issue = {7},
  pages = {075404},
  numpages = {21},
  year = {2020},
  month = {Feb},
  publisher = {American Physical Society},
  doi = {10.1103/PhysRevB.101.075404},
  url = {https://link.aps.org/doi/10.1103/PhysRevB.101.075404}
}

@article{knapp2018,
  title = {Dephasing of {Majorana-based} qubits},
  author = {Knapp, Christina and Karzig, Torsten and Lutchyn, Roman M. and Nayak, Chetan},
  journal = {Phys. Rev. B},
  volume = {97},
  issue = {12},
  pages = {125404},
  numpages = {14},
  year = {2018},
  month = {Mar},
  publisher = {American Physical Society},
  doi = {10.1103/PhysRevB.97.125404},
  url = {https://link.aps.org/doi/10.1103/PhysRevB.97.125404}
}

@article{karzig2017,
  title = {Scalable designs for quasiparticle-poisoning-protected topological quantum computation with {Majorana} zero modes},
  author = {Karzig, Torsten and Knapp, Christina and Lutchyn, Roman M. and Bonderson, Parsa and Hastings, Matthew B. and Nayak, Chetan and Alicea, Jason and Flensberg, Karsten and Plugge, Stephan and Oreg, Yuval and Marcus, Charles M. and Freedman, Michael H.},
  journal = {Phys. Rev. B},
  volume = {95},
  issue = {23},
  pages = {235305},
  numpages = {32},
  year = {2017},
  month = {Jun},
  publisher = {American Physical Society},
  doi = {10.1103/PhysRevB.95.235305},
  url = {https://link.aps.org/doi/10.1103/PhysRevB.95.235305}
}

@article{schrade2018,
  title = {Majorana Superconducting Qubit},
  author = {Schrade, Constantin and Fu, Liang},
  journal = {Phys. Rev. Lett.},
  volume = {121},
  issue = {26},
  pages = {267002},
  numpages = {6},
  year = {2018},
  month = {Dec},
  publisher = {American Physical Society},
  doi = {10.1103/PhysRevLett.121.267002},
  url = {https://link.aps.org/doi/10.1103/PhysRevLett.121.267002}
}

@article{plugge2017,
  title={Majorana box qubits},
  author={Plugge, Stephan and Rasmussen, Asbj{\o}rn and Egger, Reinhold and Flensberg, Karsten},
  journal={New Journal of Physics},
  volume={19},
  number={1},
  pages={012001},
  year={2017},
  doi={10.1088/1367-2630/aa54e1},
  publisher={IOP Publishing}
}

@article{fu2010,
  title = {Electron Teleportation via {Majorana} Bound States in a Mesoscopic Superconductor},
  author = {Fu, Liang},
  journal = {Phys. Rev. Lett.},
  volume = {104},
  issue = {5},
  pages = {056402},
  numpages = {4},
  year = {2010},
  month = {Feb},
  publisher = {American Physical Society},
  doi = {10.1103/PhysRevLett.104.056402},
  url = {https://link.aps.org/doi/10.1103/PhysRevLett.104.056402}
}

@article{nathan2020,
  title = {Universal {Lindblad} equation for open quantum systems},
  author = {Nathan, Frederik and Rudner, Mark S.},
  journal = {Phys. Rev. B},
  volume = {102},
  issue = {11},
  pages = {115109},
  numpages = {24},
  year = {2020},
  month = {Sep},
  publisher = {American Physical Society},
  doi = {10.1103/PhysRevB.102.115109},
  url = {https://link.aps.org/doi/10.1103/PhysRevB.102.115109}
}

\end{document}